\documentclass[preprint,longabstract]{aastex}
\usepackage{amsmath,mathrsfs}
\usepackage{amsfonts}
\usepackage{graphics}
\usepackage{graphicx}
\usepackage{epstopdf}
\usepackage{xfrac}
\usepackage{mathtools}
\usepackage{hyperref}

\newcommand{\kepler}{\emph{Kepler}}

\DeclareMathOperator*{\erfc}{erfc}
\newcommand{\slfrac}[2]{\left.#1\middle/#2\right.} 

\newcommand{\kms}{\ensuremath{\rm km\,s^{-1}}}

\newcommand{\gcmc}{\ensuremath{\rm g\,cm^{-3}}}

\newcommand{\Kp}{\ensuremath{K\!p}}

\newcommand{\teff}{\ensuremath{T_{\rm eff}}}
\newcommand{\logg}{log {g}}
\newcommand{\vsini}{\ensuremath{\nu \sin{i}}}
\newcommand{\vt}{\ensuremath{\nu_{\rm t }}}
\newcommand{\vrot}{\ensuremath{v_{\rm rot}}}
\newcommand{\feh}{[Fe/H]}

\newcommand{\seff}{\ensuremath{S_{\rm eff}}}
\newcommand{\searth}{\ensuremath{S_{\oplus}}}

\newcommand{\rsun}{\ensuremath{R_\sun}}
\newcommand{\msun}{\ensuremath{M_\sun}}

\newcommand{\rstar}{\ensuremath{R_\star}}

\newcommand{\rhostar}{\ensuremath{\rho_\star}}

\newcommand{\rpl}{\ensuremath{R_{\rm P}}}
\newcommand{\mpl}{\ensuremath{M_{\rm P}}}

\newcommand{\teq}{\ensuremath{T_{\rm equ}}}

\newcommand{\rearth}{\ensuremath{R_{\oplus}}}
\newcommand{\mearth}{\ensuremath{M_{\oplus}}}
\newcommand{\Searth}{\ensuremath{S_{\oplus}}}
\newcommand{\msini}{\ensuremath{m \sin{i}}}

\newcommand{\ecosw}{\ensuremath{e \cos\left(\omega\right)}}
\newcommand{\esinw}{\ensuremath{e \sin\left(\omega\right)}}
\newcommand{\Tdur}{\ensuremath{T_{\rm dur}}}
\newcommand{\Tdep}{\ensuremath{T_{\rm dep}}}

\newcommand{\blender}{{\tt BLENDER}}
\newcommand{\kea}{{\it Kea}}

\newcommand{\koicur}{Kepler-452}
\newcommand{\koicurb}{Kepler-452b}
\newcommand{\koinum}{KOI-7016.01}

\newcommand{\koicurCCra}{\ensuremath{19^{\mathrm{h}}44^{\mathrm{m}}00\fs883}}
\newcommand{\koicurCCdec}{\ensuremath{44^{\circ}16'39\farcs22}}
\newcommand{\koicurCCkic}{KIC\,8311864}
\newcommand{\koicurCCkicNoKic}{8311864}
\newcommand{\koicurCCkicr}{13.367} 
\newcommand{\koicurCCkicmag}{13.43}  
  
\newcommand{\KOIrstar}{\ensuremath{1.11^{+0.15}_{-0.09}}}
\newcommand{\KOImstar}{\ensuremath{1.037^{+0.054}_{-0.047}}}
\newcommand{\KOImstarshort}{\ensuremath{1.037}}
\newcommand{\KOIageshort}{\ensuremath{6}}
\newcommand{\KOIrhostar}{\ensuremath{0.84^{+0.40}_{-0.19}}}
\newcommand{\KOIteff}{\ensuremath{5757\pm{85}}}	%
\newcommand{\KOIfeh}{\ensuremath{0.21\pm{0.09}}}	%
\newcommand{\KOIfehshort}{\ensuremath{0.21}}	%
\newcommand{\KOIlogg}{\ensuremath{4.32\pm{0.09}}}	
\newcommand{\KOIRp}{\ensuremath{1.63^{+0.23}_{-0.20}}}
\newcommand{\KOIRpshort}{\ensuremath{1.6}}
\newcommand{\KOIperiod}{\ensuremath{384.843^{+0.007}_{-0.012}}}
\newcommand{\KOIperiodshort}{\ensuremath{384.84}}
\newcommand{\KOIepoch}{\ensuremath{314.985^{+0.015}_{-0.019}}}
\newcommand{\KOIrpRstar}{\ensuremath{0.0128^{+0.0013}_{-0.0006}}}%
\newcommand{\KOIa}{\ensuremath{1.046^{+0.019}_{-0.015}}}
\newcommand{\KOIashort}{\ensuremath{1.05}}
\newcommand{\KOIdepth}{\ensuremath{199^{+18}_{-21}}}
\newcommand{\KOIdepthshort}{\ensuremath{199}}
\newcommand{\KOIduration}{\ensuremath{10.63^{+0.53}_{-0.60}}}
\newcommand{\KOIimpactParameter}{\ensuremath{0.69^{+0.16}_{-0.45}}}
\newcommand{\KOIinsolation}{\ensuremath{1.10^{+0.29}_{-0.22}}}
\newcommand{\KOIinsolationshort}{\ensuremath{1.1}}
\newcommand{\KOIinclination}{\ensuremath{89.806^{+0.134}_{-0.049}}}
\newcommand{\KOIecosw}{\ensuremath{0.03^{+0.75}_{-0.39}}}
\newcommand{\KOIesinw}{\ensuremath{-0.02^{+0.31}_{-0.31}}}

\newcommand{\HZprobabilitySpecMatchRecentVenus}{\ensuremath{96.8}}
\newcommand{\HZprobabilitySpecMatchRunawayGreenhouse}{\ensuremath{28.0}}
\newcommand{\HZprobabilitySPCrecentVenus}{\ensuremath{99.0}}
\newcommand{\HZprobabilitySPCrunawayGreenhouse}{\ensuremath{59.3}}

\newcommand{\rockyProbabilitySpecMatchWeiss}{\ensuremath{40}}
\newcommand{\rockyProbabilitySPCWeiss}{\ensuremath{64}}

\newcommand{\rockyProbabilitySpecMatchWolfgang}{\ensuremath{49}}
\newcommand{\rockyProbabilitySPCWolfgang}{\ensuremath{62}}

\newcommand{\earthCompositionProbabilitySpecMatchWeiss}{\ensuremath{16}}
\newcommand{\earthCompositionProbabilitySpecMatchWolfgang}{\ensuremath{22}}

\newcommand{\koiTequ}{\ensuremath{265^{+15}_{-13}}}

\newcommand{\DVRpshort}{\ensuremath{1.1}}
\newcommand{\DVperiodshort}{\ensuremath{384.846}}
\newcommand{\DVTeq}{\ensuremath{221}}
\newcommand{\DVdurationshort}{\ensuremath{10.5}}
\newcommand{\DVmes}{\ensuremath{9.7}}
\newcommand{\DVdepthshort}{\ensuremath{200}}
\newcommand{\DVoddEvenTestSignificance}{\ensuremath{89.0}}
\newcommand{\DVoddEvenTest}{\ensuremath{0.14}}
\newcommand{\DVepochTestSignificance}{\ensuremath{92}}
\newcommand{\DVepochTest}{\ensuremath{0.10}}%
\newcommand{\DVstatBootstrapFAR}{\ensuremath{2.32\times10^{-16}}}%

\shortauthors{Jenkins et al.}
\shorttitle{\koicurb}

\begin{document}

\slugcomment{Received 2015 March 3; accepted 2015 May 23; published 2015 July 23 by the Astronomical Journal}

\title{Discovery and Validation of \koicurb: A \KOIRpshort-\rearth\ Super Earth Exoplanet in the Habitable Zone of a G2 Star}

\author{
Jon~M.~Jenkins\altaffilmark{1}, 
Joseph~D.~Twicken\altaffilmark{1,2}, 
Natalie~M.~Batalha\altaffilmark{1}, 
Douglas~A.~Caldwell\altaffilmark{1,2}, 
William~D.~Cochran\altaffilmark{3}, 
Michael Endl\altaffilmark{3}, 
David~W.~Latham\altaffilmark{4}, 
Gilbert~A.~Esquerdo\altaffilmark{4}, 
Shawn Seader\altaffilmark{1,2}, 
Allyson~Bieryla\altaffilmark{4}, 
Erik Petigura\altaffilmark{5}, 
David~R.~Ciardi\altaffilmark{6}, 
Geoffrey~W.~Marcy\altaffilmark{5}, 
Howard~Isaacson\altaffilmark{5}, 
Daniel Huber\altaffilmark{7,2,8}, 
Jason~F.~Rowe\altaffilmark{1,2}, 
Guillermo Torres\altaffilmark{4}, 
Stephen~T.~Bryson\altaffilmark{1}, 
Lars Buchhave\altaffilmark{4,9},
Ivan~Ramirez\altaffilmark{3}, 
Angie Wolfgang\altaffilmark{10},
Jie~Li\altaffilmark{1,2}, 
Jennifer~R.~Campbell\altaffilmark{11}, 
Peter~Tenenbaum\altaffilmark{1,2}, 
Dwight Sanderfer\altaffilmark{1}
Christopher E. Henze\altaffilmark{1},
Joseph H. Catanzarite\altaffilmark{1,2}, 
Ronald~L.~Gilliland\altaffilmark{12},
and William~J.~Borucki\altaffilmark{1}
}
\email{Jon.Jenkins@nasa.gov}
\altaffiltext{1}{NASA Ames Research Center, Moffett Field, CA 94035, USA}
\altaffiltext{2}{SETI Institute, 189 Bernardo Avenue, Mountain View, CA 94043, USA}
\altaffiltext{3}{McDonald Observatory \& Department of Astronomy, University of Texas at Austin
 2515 Speedway, Stop 1402, Austin, TX 78712, USA}
\altaffiltext{4}{Harvard-Smithsonian Center for Astrophysics, Cambridge, MA 02138, USA}
\altaffiltext{5}{University of California, Berkeley, Berkeley, CA 94720, USA}
\altaffiltext{6}{NASA Exoplanet Science Institute/Caltech Pasadena, CA 91125, USA}
\altaffiltext{7}{Sydney Institute for Astronomy (SIfA), School of Physics, University of 
Sydney, NSW 2006, Australia}
\altaffiltext{3}{Stellar Astrophysics Centre, Department of Physics and Astronomy, 
Aarhus University, Ny Munkegade 120, DK-8000 Aarhus C, Denmark}
\altaffiltext{9}{Centre for Star and Planet Formation, Natural History Museum of
Denmark, University of Copenhagen, DK-1350 Copenhagen, Denmark.}
 \altaffiltext{10}{Department of Astronomy \& Astrophysics, UC Santa Cruz, 
 1156 High Street, Santa Cruz, CA 95064, USA}
\altaffiltext{11}{Wyle Labs/NASA Ames Research Center, Moffett Field, CA 94035, USA}
\altaffiltext{12}{Center for Exoplanets and Habitable Worlds, The Pennsylvania
State University, University Park, PA 16802, USA}

\keywords{planets and satellites: detection --- stars: fundamental parameters ---
  stars: individual (\koicur, \koicurCCkic, \koinum)}

\begin{abstract}
We report on the discovery and validation of \koicurb, a transiting planet identified by a search through the 4 years of data collected by NASA's \kepler\ Mission. This possibly rocky \KOIRp-\rearth\ planet orbits its G2 host star every \KOIperiod\ days, the longest orbital period for a small ($\rpl<2$ \rearth) transiting exoplanet to date. The likelihood that this planet has a rocky composition lies between \rockyProbabilitySpecMatchWolfgang\% and \rockyProbabilitySPCWolfgang\%. The star has an effective temperature of \KOIteff\ K and a \logg\ of \KOIlogg. At a mean orbital separation of \KOIa\,AU, this small planet is well within the optimistic habitable zone of its star (recent Venus/early Mars), experiencing only 10\% more flux than Earth receives from the Sun today, and slightly outside the conservative habitable zone (runaway greenhouse/maximum greenhouse). The star is slightly larger and older than the Sun, with a present radius of \KOIrstar\ \rsun\ and an estimated age of $\sim$\KOIageshort\,Gyr. Thus, \koicurb\ has likely always been in the habitable zone and should remain there for another $\sim$3\,Gyr. 
\end{abstract}

\section{Introduction}

The study of exoplanets began in earnest just 20 years ago with the discovery of 51 Peg b, a hot jupiter in a 4.2 day period orbit of its host star \citep{mayorAndQueloz1995}. This discovery was followed by a trickle of hot jupiters that became a torrent of giant planets over a large range of orbital periods, chiefly by means of radial velocity observations. Transit surveys were already underway as early as 1994\footnote{Curiously, the first exoplanet transit survey focused on the smallest known eclipsing binary, CM draconis, to search for circumbinary planets, an endeavor that would not succeed until the \kepler\ Mission.} \citep[see, e.g.][]{doyle1996}, and rapidly gained momentum once the hot jupiter population was established.  These surveys did not bear fruit until 2000 when the Doppler-detected planet HD 209458b \citep{mazeh2000} was seen in transit \citep{charbonneau2000}. The field has advanced at a rapid pace ever since, enabling discoveries at longer orbital periods and smaller sizes.

Today $\sim$1500 exoplanets have been found through a variety of means, including Doppler surveys, ground- and space-based transit surveys, gravitational microlensing surveys \citep[starting in 1996 with OGLE-2003-BLG-235/MOA-2003-BLG-53,][]{bennett2006}, and recently, direct imaging \citep[beginning in 2011 with HR 8799,][]{soummer2011}. As impressive as the progress has been in this nascent field of astronomy, perhaps the most exalted goal, and one that remains tantalizingly close, is the discovery of a sufficient number of (near) Earth--Sun analogs to inform estimates of the intrinsic frequency of Earth-size planets in the habitable zone of Sun-like stars.

The habitable zone is an evolving concept, here taken to be that range of distances about a star permitting liquid water to pool on the surface of a small rocky planet. For our purposes, we adopt the optimistic habitable zone as per \citet{kopparapu2013} with the insolation flux, \seff,  between that experienced by recent Venus as the inner edge, and that experienced by early Mars as the outer edge. While the exact values depend somewhat on the effective temperature of the star, the optimistic habitable zone lies within the range from $\sim$20\% to $\sim$180\% of the radiation experienced by earth today, that is from 0.2\,\searth\ to 1.8\,\searth. 

To date, nine small planets with radii less than  2.0\,\rearth\ have been found in the 4 years of \kepler\ data in or near the habitable zones of their stars. These discoveries include Kepler-62e and f \citep{kepler62}, Kepler-186f \citep{kepler186f}, a few among the multitude of planets reported by \citet{roweMultis2014}: Kepler-283c, Kepler-296e and f; and a set of small planets announced in \citet{torres2015}: Kepler-438b, Kepler-440b, and Kepler-442b. The repurposed \kepler\ Mission, dubbed K2, has also added a small planet on the inside edge of its habitable zone: K2-3d  \citep[aka EPIC 201367065d;][]{k2-3c}. Most of these small planets are beyond the reach of Doppler surveys and cannot be confirmed independently by radial velocity detections, and therefore must be validated statistically. Follow-up observations are conducted to rule out, to a very high statistical significance, scenarios in which astrophysical signals can mimic the planetary transits under investigation.

Doppler surveys have been successful for a handful of very bright (nearby), cool stars (where the habitable zone is close-in and reflex velocity amplitude is high). These possibly rocky planets ($\msini <10\,\mearth$) include GJ 163c \citep{GJ163c2013}, GJ 667 Cc  \citep{gj667CcAngeladaEscude2013,delfosse2013},  HD 40307g \citep{tuomi2013}, Kapteyn b \citep{kapteynb2014}, and GJ 832c \citep{GJ832c2014}. 

All of these are orbiting small ($\rstar<\rsun$), cool ($\teff < 5000$\,K) stars. According to the recent \kepler\ Q1--Q16 catalog \citep{Q1_Q16TCEs2015} available at the NExSCI exoplanet archive,\footnote{\url{http://exoplanetarchive.ipac.caltech.edu/cgi-bin/ExoTables/nph-exotbls?dataset=cumulative_only}} approximately 65 additional small planetary candidates have been identified orbiting in or near the habitable zones of stars over a much wider range of spectral types with effective temperatures from 2703 K to as high as 6640 K. All of these candidates need to be vetted thoroughly, given their importance to the prospect of reaching the primary \kepler\ Mission objective: to determine the frequency and distribution of terrestrial planets in the habitable zones of Sun-like stars.

Here we present the detection and statistical validation of \koicurb, a possibly rocky \KOIRpshort-\rearth\ planet orbiting in the habitable zone of its G2 host star \koicurCCkic\ every \KOIperiodshort\ days, the longest orbital period for a small ($\rpl < 2$\,\rearth) transiting exoplanet discovered to date. In this paper, we apply the \blender\ analysis technique (see Section~\ref{s:blender}) to validate the planetary nature of \koicurb\ , but note that similar approaches have been developed by others and are in use in transit surveys \citep[see, e.g.][]{morton2012,Barclay13,diaz2014}.

The paper is organized as follows: Section~\ref{s:discovery} describes the initial discovery of the transit signature of this super Earth in 2014 May. Section~\ref{s:photometry} presents the photometry for this transiting planet, describing the discovery of the transit signature in the data, and section~\ref{s:DV} describes the results of several tests performed in the \kepler\ science pipeline to help validate the planetary nature of the photometric signature. Section~\ref{s:spectroscopy} describes the spectroscopic measurements obtained for this host star, and section~\ref{s:hostStarProperties} presents the stellar properties obtained for \koicurCCkic. The analytical approach used to fit the light curve and determine the planetary properties is described in Section~\ref{s:planet properties}. Section~\ref{s:adaptiveoptics} details the adaptive optics observations made to search for close companions. Section~\ref{s:blender} describes the results of a full \blender\ validation analysis, which indicate that the planet hypothesis is favored at an odds ratio of 424:1, providing high confidence that this is, indeed, a planetary signature. Section~\ref{s:habitabilityandcomposition} discusses the history of \koicurb\ in the context of the evolving habitable zone of its host star. Section~\ref{s:conclusions} summarizes the results. The appendix presents a mathematical development of the bootstrap diagnostic calculation.

\section{Discovery}\label{s:discovery}
\koicurb\ was discovered in a test run of the \kepler\ Science Operations Center (SOC) 9.2 codebase in 2014 May when one of us (J. Twicken) inspected the planet search pipeline results to assess performance of an enhanced pipeline codebase for small, cool planets.  According to the Data Validation pipeline module, the transit signature of this object featured four \DVdurationshort hr, \DVdepthshort-ppm deep transits spaced \DVperiodshort\ days apart, a radius of \DVRpshort\,\rearth, and an equilibrium temperature of \DVTeq\ K assuming an albedo of 0.3 and full redistribution of heat. None of the Data Validation diagnostics definitively rejected the planetary nature of this signature, and while both the radius and the equilibrium temperature proved to be significantly larger once better stellar parameters were obtained, this was clearly a highly interesting threshold crossing event (TCE).

This transit signature was not identified in the previous search through all 4 years of \kepler\ data with the SOC 9.1 codebase conducted in 2013 August \citep{tenenbaum2014}, most likely due to overly aggressive consistency checks put into place to reduce the large number of false alarms by instrumental effects \citep{seaderVetos2013}. 
After all 4 years of \kepler\ data had been re-processed with the deployed SOC 9.2 codebase, \koicurb's transit signature was re-identified in the recent run of the planet search pipeline in 2014 November \citep{seaderTCEs2015}. The re-processed light curves and calibrated pixels have been archived at the Mikulski Archive for Space Telescopes (MAST) as data release 24, and the transit search results have been archived at NASA's Exoplanet Science Institute's exoplanet archive.\footnote{Available at \url{http://exoplanetarchive.ipac.caltech.edu/cgi-bin/TblView/nph-tblView?app=ExoTbls&config=q1_q17_dr24_tce}.} This planet has been designated as \koinum\ and appears in the online KOI catalog table on NExScI with the first delivery of the Q1 -- Q17 KOIs. We expect that several more terrestrial planetary candidates will be identified when all the new SOC 9.2 TCEs are fully vetted. 

This is the first planetary signature detected for \koicurCCkic, which is noteworthy, given the long orbital period and the high fraction of multiple transiting planet systems identified by \kepler\ \citep[see e.g.,][]{roweMultis2014}. If there are planets in orbits interior to that of \koicurb, either they are inclined so as not to present transits or they are too small to detect against the photometric noise for this 13.4 mag star (which averaged $\sim$40\,ppm at 10.5 hr timescales across the 4 year dataset). 

The stellar parameters for this star at the time of the test run included an effective temperature of 5578\,K but a radius of only 0.79\,\rsun, yielding a planet with a correspondingly small radius of 1.1\,\rearth\ at a mean separation of 0.99\,AU. The small stellar radius was due to an anomalously high \logg\ of 4.58.\footnote{The original KIC value for the \logg\ of 4.99 \citep{brown2011} was corrected by moving the stellar properties to the nearest Yonsei-Yale isochrone, giving a still somewhat high value of 4.58.} Given that the star was likely to be significantly larger than 0.79\,\rsun\ on account of its temperature, we sought reconnaissance  spectra in order to better pin down the stellar parameters and eventually obtained a spectrum with the Keck HIRES spectrometer. These measurements indicated that the surface gravity of the star was overstated by the KIC, and thus imply that the planet is $\sim$45\% larger than it appeared in the Data Validation summary report produced in the 2014 May test run of the \kepler\ pipeline.

\section{\emph{Kepler} Photometry} \label{s:photometry}

The \emph{Kepler} telescope obtained nearly continuous photometry in a 115\,deg$^2$ field of view (FOV) near Cygnus and Lyra for nearly 4 years  \citep{borucki2010}. During this period (2009 May 13  through 2013 May 11), \kepler\ observed 111,800 stars nearly continuously and altogether observed a total of more than 200,000 stars over its operational mission. The \kepler\ observations were broken nominally into 3 month (93 day) segments, dubbed quarters, due to the need to rotate the spacecraft by $90^\circ$ to keep its solar arrays pointed toward the Sun and its sunshade properly oriented \citep{koch2010}.

The star \koicur\ (Kepler Input Catalog (KIC) = \koicurCCkicNoKic, $\alpha = \koicurCCra$, $\delta = \koicurCCdec$, J2000, KIC~$r = \koicurCCkicr$\,mag) was observed during each of the 17 quarters (Q1, Q2, $\ldots$, Q17) returned by \kepler.\footnote{A set of $\sim$53,000 stars including \koicurCCkic\ were observed for a 10 day period at the end of the commissioning period for \kepler\ that is called Q0. This data set is not used in the search for planets or for constructing pipeline diagnostics in the Data Validation module, and is not used in this analysis except where otherwise noted.} The first and last quarters were very much abbreviated compared to the others: Q1 lasted 34 days and took place immediately after commissioning was complete \citep{haas2010}, while Q17 lasted only 32.4 days, being truncated by the failure of a second reaction wheel, thus ending the data collection phase of the \kepler\ Mission. 

\begin{figure*}
\resizebox{\hsize}{!}{\includegraphics{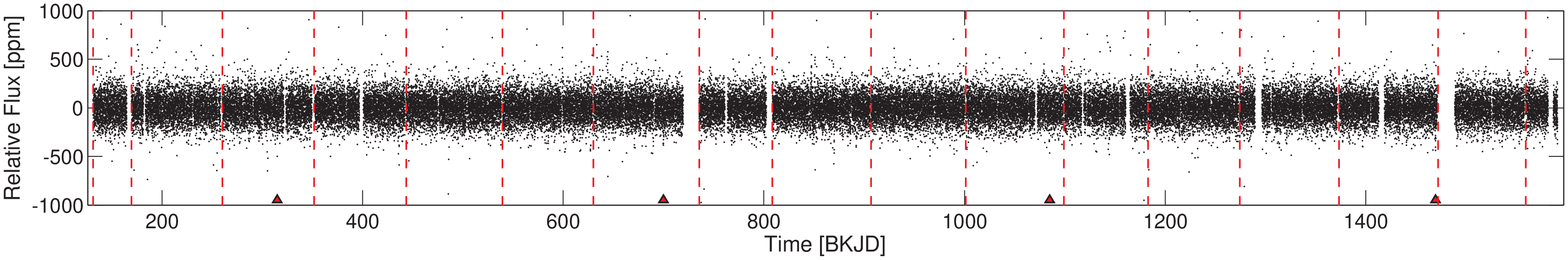}}%
\caption{
Detrended flux time series for \koicur\ versus barycentric-corrected \kepler\ Julian day (BKJD - 2,454,833.0). The triangle symbols indicate the locations of each of the four transits which occur in quarters 3, 7, 11 and 15. All 4 years of \kepler\ data are plotted here with vertical dashed red lines indicating the boundaries between consecutive quarters. The detrending was accomplished by subtracting the result of passing the systematic error-corrected light curve through a moving median filter with a window width of five times the $\sim$\DVdurationshort hr pulse duration. The light curve residuals are remarkably uniform across all 17 quarterly data segments, showing no evidence of electronic image artifacts which are responsible for a large number of long period false positive transit-like features in the \kepler\ light curves.
\label{fig:detrendedlightcurve}}
\end{figure*}

Figure~\ref{fig:detrendedlightcurve} shows all 17 quarters of the systematic error-corrected flux time series data for this star, with the locations of the transits identified by triangle markers.\footnote{All \kepler\ photometric data products can be found at the MAST archive at http://archive.stsci.edu/kepler.} The first transit takes place during Q3 at a barycentric-corrected \kepler\ Julian date (BKJD)\footnote{Barycentric-corrected \kepler\ Julian date (BKJD) is defined as BJD$-$2,454,833.0.}  of \KOIepoch, and three additional transits occur every \KOIperiod\ days, one each in Q7, Q11 and Q15. The transits each last approximately \DVdurationshort\ hours and are \DVdepthshort\ ppm deep. The scatter in the photometric data in Figure~\ref{fig:detrendedlightcurve} is quite uniform in all quarters and there is no obvious evidence of the electronic image artifacts that plague a subset of the \kepler\ target stars as they rotate through the handful of affected CCD readout channels, resulting in a  large number of false alarm TCEs \citep{vancleveandcaldwell2009,tenenbaum2013,tenenbaum2014}. Indeed, this star does not fall on any of the CCD channels associated with these image artifacts in any observing season.

\begin{figure*}[ht!]
\resizebox{\hsize}{!}{\includegraphics{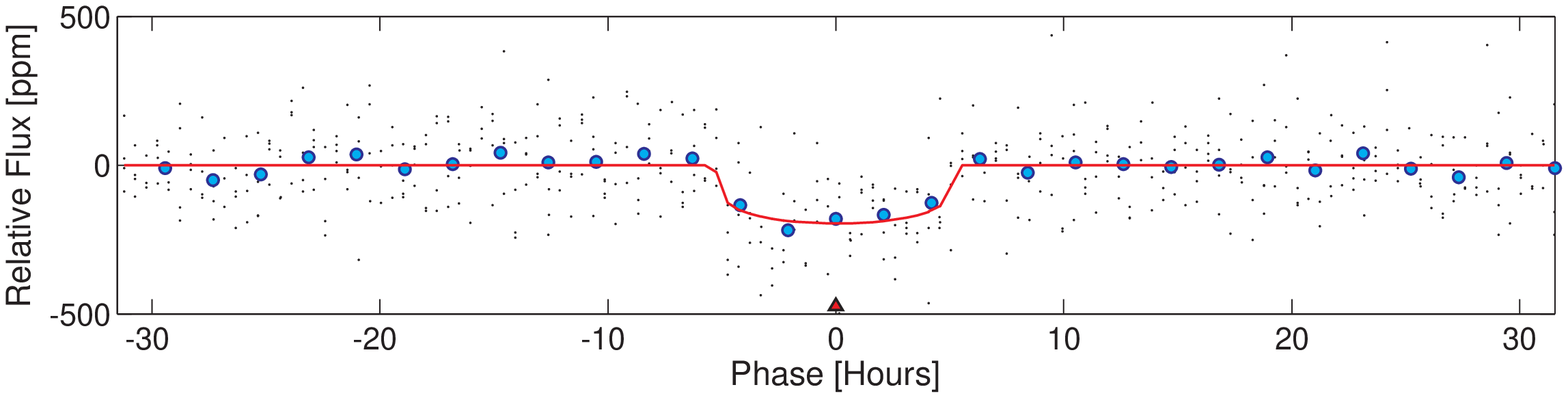}}
\caption{
Phase-folded, detrended flux time series for \koicur\ versus phase in hours. The flux time series plotted here is the same as that of Figure~\ref{fig:detrendedlightcurve}, except that it has been phase folded at the period of the planet, \KOIperiodshort\ days. The triangle symbol indicates the location of the four transits at zero phase. The full resolution data are represented by black dots. The cyan filled circles represent the results of binning the data in phase with a bin width of $\sfrac{1}{5}$ of the DV-fitted duration of the transit,  \DVdurationshort\,hr. 
\label{fig:phasefoldedzoomedlightcurve}}
\end{figure*}

The planet was detected at the $\DVmes \,\sigma$ level, well above the $7.1\, \sigma$ threshold and passed the other consistency checks exercised by the Transiting Planet Search (TPS) component of the SOC pipeline \citep{jenkins2002, jenkinsTPS2010, seaderVetos2013, tenenbaum2013, tenenbaum2014}. The light curve with its putative transit signature was  fitted with a \citet{MandelAndAgol2002} limb-darkened transit waveform and subjected to a suite of diagnostic tests by the Data Validation module \citep{dvFitSpie2010,dvSpie2010}. Figure~\ref{fig:phasefoldedzoomedlightcurve} shows the phase-folded,  detrended light curve, zoomed on the transit epoch. The individual transits are significant at the $\sim$4.75$\,\sigma$ level, so that each can be credibly seen by eye in the long cadence (LC) data at a 29.4 minute resolution. The DV diagnostics appeared to be consistent with the planetary hypothesis, so we coordinated follow-up observations to help refine the stellar (and hence planetary) properties and to further bolster the case for the planetary nature of this object.

The following section describes the results of the validation tests performed in DV.

\section{Data Validation Tests}\label{s:DV}
This section describes the most salient validation tests performed inside of the Data Validation component of the SOC pipeline. The DV report resulting from the official SOC 9.2 processing is available at  \url{http://exoplanetarchive.ipac.caltech.edu/data/KeplerData/008/008311/008311864/dv/kplr008311864-20141002224145_dvr.pdf}.\footnote{The reprocessed light curves are available at \url{http://archive.stsci.edu/kepler/} and the release notes can be found at \url{http://archive.stsci.edu/kepler/release_notes/release_notes24/KSCI-19064-001DRN24.pdf}.}

\subsection{Discriminating Against Background Eclipsing Binaries}\label{DVeclipsingbinaries}
\begin{figure}[ht]
\resizebox{\hsize}{!}{\includegraphics{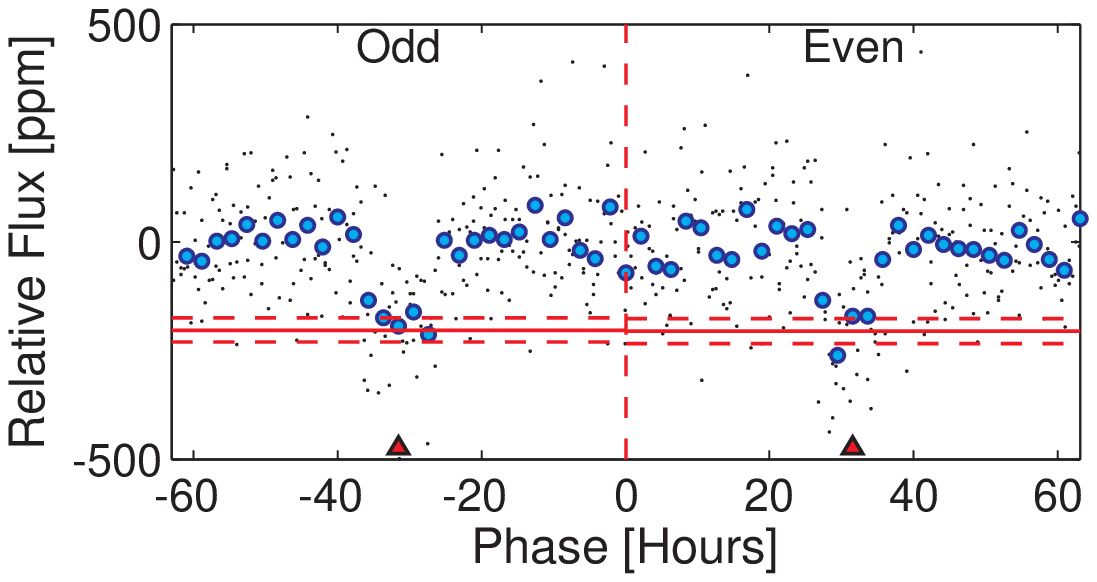}}
\caption{
Phase-folded, detrended flux time series for \koicur\ versus phase in hours as in Fig.~\ref{fig:phasefoldedzoomedlightcurve}, with the data for the even transits separated from that of the odd transits. The full resolution data are represented by black dots. The cyan fllled circles represent the results of binning the data in phase with a bin width of $\sfrac{1}{5}$ of the DV-fitted duration of the transit,  \DVdurationshort\ hours. The red horizontal lines are the fitted depths for the odd transits (left) and the even transits (right) and the dashed horizontal lines represent the $\pm 1\, \sigma$ error bars on the fitted depths.
\label{fig:oddEvenTest}}
\end{figure}

Data validation applies several tests for the presence of background eclipsing binaries. The first is the odd/even test: DV fits a  limb-darkened transit model to the odd transits and an independent model to the even transits, and then compares the depth fitted to the odd transits to that fitted to the even transits. This results in a test statistic of $\DVoddEvenTest\, \sigma$ with a corresponding significance of \DVoddEvenTestSignificance \%.\footnote{The significance of these statistical tests is relative to the null hypothesis. That is, a significance near 1 indicates a null result (not an eclipsing binary) while a significance near zero indicates that the alternative hypothesis is highly likely.} In addition, DV examines the ratio of the temporal offsets between the odd and even transits compared to the fitted orbital period, the so-called ``epoch'' test, giving a result of $\DVepochTest\, \sigma$ and a corresponding significance of \DVepochTestSignificance \%. Figure~\ref{fig:oddEvenTest} illustrates the odd/even test by plotting the light curve data phase folded separately for the odd and even transits. No other TCEs were generated for this star and no evidence was seen for weak secondary eclipses. Based on these tests, we see no evidence for systematic differences in the odd/even transit depths and durations which often result in the case of background eclipsing binaries, and the transit times are consistent with a strict periodic spacing.

\subsection{Statistical Bootstrap Test}\label{ss:bootstrap}
In order to better determine the statistical significance of the transit sequence, we examined the bootstrap distribution for the population of null statistics for the out-of-transit data \citep[][J. Jenkins \& S. Seader, in preparation]{jenkinsetal2002}. This test relaxes the assumption that the observation noise is broadband colored Gaussian noise to establish a robust significance level for the detection statistic of the planet candidate. The appendix contains a mathematical derivation of the statistical approach and a brief description of its implementation.

\begin{figure}[ht!]
\resizebox{\hsize}{!}{\includegraphics{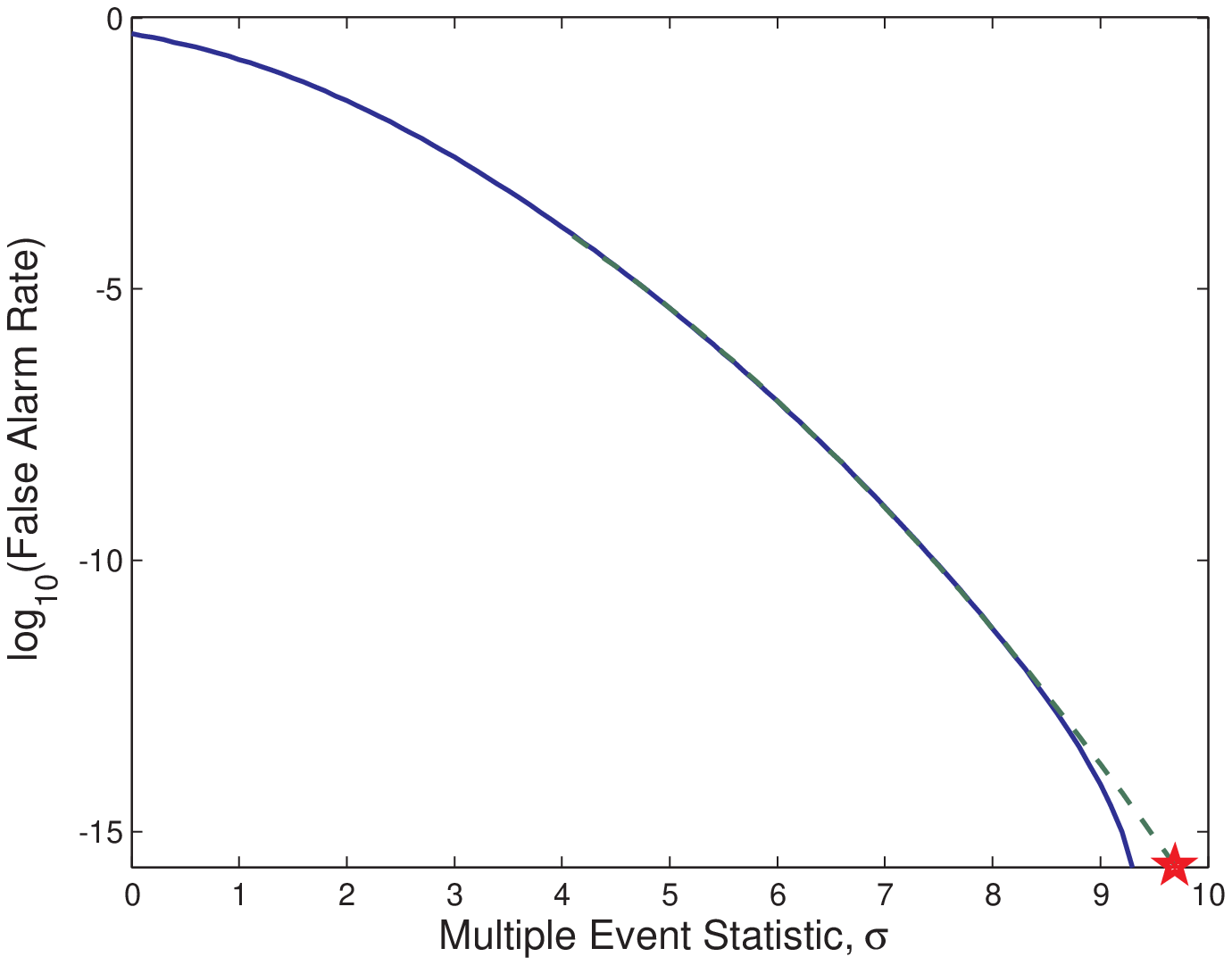}}
\caption{
Bootstrap false alarm rate (FAR) versus maximum multiple event statistic for the light curve for \koicur\ (blue curve), and that expected for ideal Gaussian noise fitted to the empirical FAR between $1\times10^{-4}$ and $1\times10^{-13}$ (dashed green curve). The Gaussian fit allows us to extrapolate the FAR out to the maximum $MES$ provided by the planetary signature as the highest $MES$ probed by the bootstrap distribution falls short of the 9.7 $\sigma$ $MES$ of this transiting planet in the \kepler\ data. The red star indicates the estimated bootstrap FAR obtained from the Gaussian fit. 
\label{fig:bootstrap}}
\end{figure}

The false alarm rate is estimated using a bootstrap statistical analysis of the out-of-transit data for this star. The single event statistics time series, representing the likelihood of a transit of the fitted duration as a function of time, are drawn at random (with replacement) to construct artificial multiple event statistics corresponding to the same number of transits that occurred for the TCE in question. The distribution of all such bootstrap samples provides a means to assess the statistical significance of the maximum multiple event statistic, the $maxMES$, that resulted in the identification of the TCE in the first place. This methodology relaxes the assumption of perfection on the part of the adaptive, power spectral density estimator in TPS that informs the pre-whitener, allowing for the detection signal-to-noise ratio (S/N) to be evaluated against the distribution of null statistics drawn directly from the observation noise of the light curve. The bootstrap false alarm rate for the transit signature of \koicurb\ is securely established as statistically significant by this test for the $\DVmes\,\sigma$ detection of this planet, at a level of \DVstatBootstrapFAR. This figure is well below that required by the \kepler\ experiment, $\sim1\times10^{-12}$, to limit the number of false positives to no more than one over the full mission. Figure~\ref{fig:bootstrap} illustrates the bootstrap test for this planet.

\subsection{Centroid Analysis}\label{ss:centroidanalysis}

Data validation performs two types of pixel-level centroid analysis \citep{brysonFalsePositives2013} to determine the location of the transit source relative to \koicurCCkic.  This analysis constrains the likelihood that transits observed in a light curve may be due to background eclipsing binaries by providing a ``confusion radius,'' inside which a background transit source cannot be distinguished from the target star.  Sources outside this confusion radius are ruled out as possible causes of the transit signal.  This radius is passed to the \blender\ analysis of Section~\ref{s:blender}.

The first analysis technique uses the change in the photocenter (centroid shift) of the flux on pixels when the transit occurs to infer the location of the source of the transit signal. The row and column of the photocenter of pixels collected for this target are measured for every LC in which the target was observed. These row and column time series are then fitted to the transit model (normalized to unit depth) derived from the light curve. The fitted amplitudes of the row and column fits provide the apparent centroid shift on the sky.  The use of every observed LC in this fit allows a very high precision measurement of this centroid shift.  Dividing this centroid shift by the transit depth gives an estimate for the transit source location relative to the target star.  For \koicurCCkic, the source offset  was estimated as 1\farcs832 at $1.63\,\sigma$, consistent with the transit source being co-located with \koicurCCkic. This estimate, however, tends to be biased so as to overestimate the source offset relative to the mean photocenter of the pixels. It also tends to be more vulnerable to noise, instrumental systematics, and other flux variations in the photometric mask than estimates obtained via the second technique.  

The second method consists of performing difference image analysis on in-transit pixel data compared with neighboring out-of-transit pixel data in each quarter. The average of the in-transit images is subtracted from the averaged out-of-transit images. When flux variation in the photometric mask other than the transit is negligible, the transit source appears as a star in the difference image. Fitting the quarterly difference and out-of-transit images with the \kepler\ point-spread function (PSF) then provides a direct measurement of the offset of the transit source from \koicurCCkic\ per quarter.  The quarterly offsets are combined via robust averaging to provide an estimate of transit source offset.  The difference image analysis generally tends to be less vulnerable to noise and other flux variation than the brightness weighted centroid results.  For shallow transits which occur only once per quarter, such as \koicurb, however, instrumental systematics can result in very poor (un-star-like) difference images whose PSF fit will result in significantly biased position estimates.  When the systematics are sufficiently strong, the fit can fail to converge.  

\begin{figure}[ht]
\resizebox{\hsize}{!}{
\includegraphics{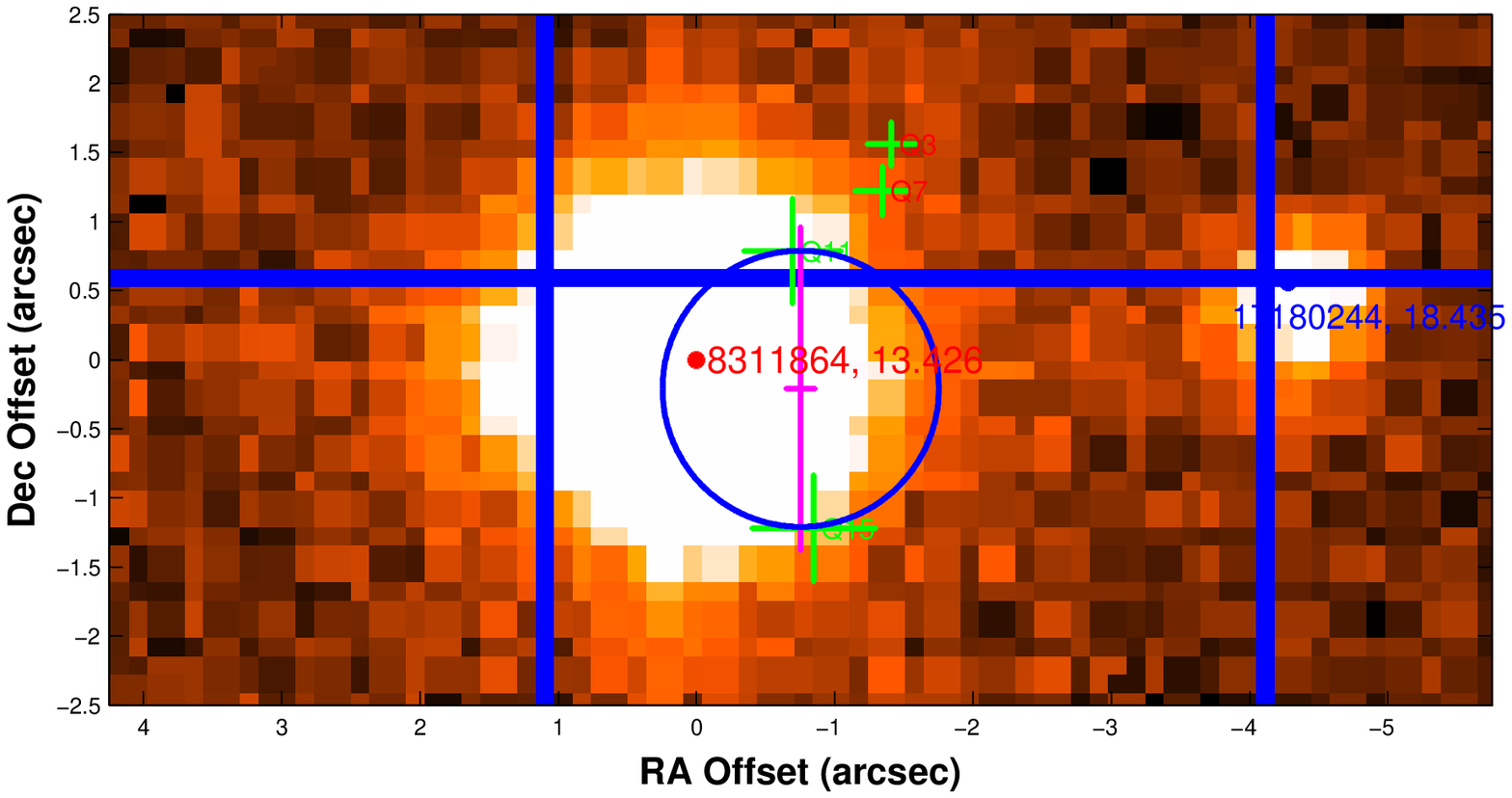}}
\caption{Difference image centroiding results for \koicurb\ with a $J$-band infrared image from the UK Infrared Telescope Survey \citep[UKIRT;][]{lawrence2007} overlaid. There is a faint, 18$^\mathrm{th}$ mag star offset by $\sim$4\arcsec\ to the west of \koicurCCkic, but neither the difference image centroids in this figure, nor the centroid shift analysis provide any evidence suggesting that this star is the source of the transit-like features. The four green crosses indicate the difference image centroids for each of the four transits of this super Earth. The blue circle delimits the 3-$\sigma$ confusion radius, which contains the target star, \koicurCCkic, indicating that the target star is the most likely source of the transit features in the \kepler\ light curve, to the limits of the resolution of the UKIRT images.
\label{fig:diffImages}}
\end{figure}

For \koicurCCkic, all four quarters containing transits had successful fits, shown as the green crosses in Figure~\ref{fig:diffImages}.  Two of the quarters, however, had very poor, non-star-like difference images and resulted in the two locations to the northwest of \koicurCCkic.  The other two quarters had good, star-like images, resulting in offsets that straddle \koicurCCkic, slightly offset to the west.  Because the two non-star-like difference images are not validly fit by the \kepler\ PSF (the difference images in these quarters do not contain anything that looks like a star), we exclude them from our analysis.  The remaining two quarters give a transit source offset from \koicurCCkic\ of $0\farcs782\pm0.33$, or $2.35\,\sigma$.  This measurement formally has a $3\,\sigma$ confusion radius of 1\farcs0.  

To address the concern that a measurement using only two data points can be misleading, we estimate that a Gaussian uncertainty of 0\farcs50 would result in the observed uncertainty of 0\farcs33 half the time.  We therefore conservatively set the 3-$\sigma$ confusion radius to 1\farcs5.  Potential transit sources 
outside this 1\farcs5 radius are ruled out. Possible unknown sources inside 1\farcs5 are addressed in the
\blender\ analysis of Section~\ref{s:blender}.

\subsection{Contamination by Other Sources}\label{ss:contamination}
In addition to the tests conducted by the pipeline module Data Validation on the \kepler\ data themselves, we also performed manual checks against contamination by other astrophysical sources in the \kepler\ FOV. The two principal sources of contamination are (1) optical ghosting from eclipsing binaries well offset from the target star due to reflections off of optical elements in the \kepler\ telescope, and (2) video crosstalk between CCD channels that are read out simultaneously.

Optical ghosting, or direct PSF contamination, was the subject of a study by \citet{coughlin2014} identifying 685 TCEs in the Q1--Q12 vetting process as false positives due to these and other effects. To determine whether optical ghosting was an issue, we looked for period-epoch collisions between \koicurb\ and the TCEs identified in both the 2012 August search through Q1--Q16 \citep{tenenbaum2013} and those from the recent 2014 September search through Q1--Q17 \citep{seaderTCEs2015}. None were found. Nor were any TCEs found either at one half or one third of the orbital period of \koicurb\ with transit epochs matching the epochs of \koicurb. 

In addition, we queried the general catalogue of variable stars (GCVS) \citep{GCVS} for variable stars with periods within 10 hours of \koicurb. Only two stars in the GCVS had periods close to \KOIperiodshort\ days that were within $\sim$6$^\circ$ of \koicurCCkic. Both of these are pulsating variables, however, not eclipsing binaries, and therefore cannot explain the transits of \koicurb. One of them is FX CYG aka KIC 4671784, which was observed in LC mode throughout \kepler's mission. The other variable star is V1299 Cyg which is more than 1$^\circ$ from the \kepler\ FOV and thus cannot be the source of any contamination. No variable stars were found in the GCVS in or near \kepler's FOV with periods near one half or one third that of \koicurb. The likelihood that optical ghosting could explain the transits of \koicurCCkic\ is further diminished by the fact that most optical ghosts tend to create multiple ``children'' signals in the \kepler\ light curves as the ghost images are defocussed or severely scattered.
 	
Video crosstalk occurs when analog voltage signals in one video circuit impress themselves onto other video circuits in the readout electronics boards \citep[][p. 71]{vancleveandcaldwell2009}, contaminating the other CCD readout values. This represents an additive noise source and the contamination is an attenuated version of the original signal, but can either be positive or negative, with a typical value of $\sim$4400 for the attenuation factor (or a gain factor of $2.3\times10^{-3}$). While video crosstalk does occur between channels read out at the same time on different CCD modules, the attenuation is so high that the only appreciable video crosstalk occurs within each pair of CCDs mounted together on the same module. To rule out contamination by video crosstalk, we examined the quarter 3 month 2 full frame image (FFI) \url{kplr2009322233047_ffi-cal.fits} obtained on UT 2009 November 18 for evidence of bright stars read out on the same pixels as \koicurCCkic\ on the adjacent CCD channels.\footnote{FFIs are available at \url{http://archive.stsci.edu/kepler/ffi/search.php}.}  

For a target star of brightness $B$ experiencing video crosstalk from a perfectly situated eclipsing binary (or transiting giant planet) of brightness $b$ and an eclipse (or transit) depth of $\slfrac{\delta b}{b}$ on an adjacent CCD readout channel, the apparent transit depth exhibited by the  target star is given by
\begin{equation}
\label{eq:videocrosstalk}
\begin{split}
\Tdep \times 10^{-6} & = \frac{\alpha \, \delta b}{B+\alpha b}=\frac{\alpha\,\slfrac{\delta b}{b}}{\slfrac{B}{b}+\alpha}\\
& \approx \frac{\alpha\,\slfrac{\delta b}{b}}{\slfrac{B}{b}} = \alpha \frac{\delta b}{b} \frac{b}{B},
\end{split}
\end{equation}
where the second line arises for the case of interest here as $\alpha \ll 1$.

All four transits of \koicurCCkic\ occur on CCD module/output 7.3, as the orbital period is close to one year and the first transit occurs sufficiently early in Q3 so that the subsequent three transits take place when the \kepler\ photometer is in the same seasonal orientation. Thus, only stars on the adjacent channels 7.1, 7.2, and 7.4 could produce video crosstalk contamination for \koicurb. Of these, only channel 7.1 has a readily visible star ($\alpha = 19^{\mathrm{h}}44^{\mathrm{m}}27^{\mathrm{s}}.95$, $\delta = 42^{\circ}46'13''.55$) near the same row/column coordinates as for \koicurCCkic\ on channel 7.3. Its peak pixel actually is just 1 row below the optimal aperture for Q3, so only flux from the wing of the PSF could be getting into the aperture of \koicurCCkic. No \kepler\ light curves exist for this star since it was not selected as a target during the 4 year \kepler\ Mission.

This star on channel 7.1 is about 3 mag fainter than \koicurCCkic\ and the video crosstalk coefficient for channel 7.1 on channel 7.3 is 2.3820$\times10^{-5}$. Hence, a full eclipse (50\% drop in brightness) on this star on channel 7.1 would produce an apparent transit depth of $\sim$0.75\,ppm. A 1\% giant planet transit on this star would impute an even smaller apparent transit depth of $\sim$0.015\,ppm. We can thus eliminate this star as a potential source of confusion as it cannot credibly cause a transit-like feature deep enough to explain the \KOIdepthshort\,ppm transits exhibited by \koicurb. 

Channels 7.2 and 7.4 are also incapable of producing a false positive here through video crosstalk. Channel 7.2 has a negative crosstalk coefficient with 7.3 (-1.9759$\times10^{-4}$) and 7.4 has a crosstalk coefficient of 1.276$\times10^{-4}$, but no apparent star in the  pixels corresponding to the aperture of \koicurCCkic.

We conclude that contamination through either optical ghosting or video crosstalk is highly unlikely for \koicurb.

\section{Spectroscopy}\label{s:spectroscopy}
Due to the curiously small stellar radius of \koicurCCkic\ for the effective temperature available at the time of the 2014 May planet search, it was clear that reconnaissance spectroscopy was called for in order to obtain a reasonable interpretation of the nature of the source of the transit-like features in the \kepler\ light curve. We obtained one spectrum from the  McDonald Observatory in 2014 May and two additional spectra from the Fred L. Whipple Observatory in 2014 June. To improve the stellar parameters derived from spectroscopy, we obtained a spectrum from the Keck I observatory with the HIRES instrument in 2014 July. The HIRES data were analyzed with three independent analysis packages in order to better understand the effects of systematics in the software codes and to determine the most reasonable stellar parameters to use for interpreting the photometry in this system.
\subsection{Reconnaisance Spectroscopy with the Tull spectrograph at the McDonald Observatory 2.7 m Telescope}

We obtained a reconnaissance spectrum for \koicurCCkic\ using the Tull Coud\'e spectrograph \citep{tull1995} at the Harlan J. Smith 2.7 m telescope at McDonald Observatory. We used the standard instrumental setup that covers the entire optical spectrum and uses a 1\farcs2 slit that yields a resolving power of $R$ = 60,000 with 2-pixel sampling. We observed the star in the night of UT 2014, 18 May. The exposure time was 1720 seconds and the resulting spectrum has a S/N of around 31:1 (per resolution element) at 5650\AA. We retrieved stellar parameters using our new spectral fitting tool \kea. \kea\ was specifically developed to determine reliable stellar parameters from the lower S/N reconnaissance spectra of our \kepler\ follow-up program at McDonald Observatory. 

We used the MOOG \citep{sneden1973} stellar spectral code in its spectral synthesis mode with the \citet{kurucz1993} stellar atmospheric models to construct a grid of synthetic stellar spectra.  We used atomic line parameters obtained from the VALD (Vienna Atomic Line Database). We included molecular opacities for MgH \citep{mgh2013}, TiO \citep{plez1998} and CN \citep{sneden2014}. The spectral grid covers a range of effective temperature $T_{\rm eff}$ from 3,500 to 10,000 K in 100 K steps, a stellar metallicity of [Fe/H] from -1.0 to +0.5 dex in 0.25 dex steps, and a surface gravity log(g) of 1.0 to 5.0\,dex in steps of 0.25\,dex. We used the MOOG ``weedout'' feature to eliminate atomic and molecular lines in the line list with a ratio of line to continuum opacity of less than 0.001.

\kea\ compares 12 orders of the continuum-normalized Tull spectrum with the synthetic grid using $\chi^{2}$ statistics. For this comparison we convolve the synthetic spectra with a Gaussian-shaped PSF to assure the same spectral resolution of model and data. We also rotationally broaden the model spectra to find a best match for the (projected) \vrot\ of the star. The final stellar parameters are obtained from the mean and standard deviation of the best-fit models of the spectral orders that \kea\ is using. A more detailed description of this method will be presented in a forthcoming publication (M. Endl \& W. Cochran, in preparation). From the KIC 8311864 reconnaissance spectrum, we determined an effective temperature of $\teff = 5650\pm108$\,K, a metallicity of $\feh = 0.21\pm 0.07$\,dex, a surface gravity of $\logg =4.45\pm0.16$, and a rotational broadening of $\vrot\ = 4.30\pm0.5$\,km\,s$^{-1}$ for \koicurCCkic.    

\subsection{Reconnaisance Spectra Obtained from TRES}\label{sss:TRESspectra}
We obtained two spectra using the Tillinghast Reflector Echelle Spectrograph \citep[TRES;][]{furesz:2008}, on the 1.5\,m telescope at the Fred Lawrence Whipple Observatory (FLWO) on Mt. Hopkins, Arizona, which has a resolving power of $R=44,000$ and a wavelength coverage of $\sim$3900--9100\,\AA. The spectra were taken on UT 2014, June 09 and UT 2014, June 14 and were extracted as described by 
\citet{buchhave:2010}. 

We used the Spectral Parameter Classification (SPC) \citep{buchhave:2012} technique to determine stellar parameters. SPC cross correlates an observed spectrum against a grid of synthetic spectra using the correlation peak heights to determine the best fit parameters. The synthetic spectra are based on Kurucz model atmospheres \citep{kurucz1993}. We allowed \teff, \logg, \feh, and \vsini\  to float as free parameters. The S/N per resolution element on the TRES spectra were slightly lower than usual when running SPC so these results should be viewed with caution. 
A weighted average of the results from the  two spectra gave \teff\ = $5751 \pm 55$\,K, \logg\ = $4.43\pm0.10$, \feh\ = $0.40\pm0.08$, and \vsini\ = $4.2\pm0.5$\,\kms.  The maximum height of the normalized cross correlation function (CCF) was 0.916 and the combined S/N of the spectra was 23.5.

\subsection{HIRES Spectroscopy}\label{ss:HIRES}

On UT 2014, July 3, we acquired a high resolution spectrum of \koicurCCkic\ with the Keck I Telescope and HIRES spectrometer. We used the standard setup of the California Planet Search \citep{howard2010} without the iodine cell, resulting in a spectral resolution of ~60,000, and S/N, at 5500\,\AA, of 40 per pixel. 

SpecMatch compares a target stellar spectrum to a library of 800 spectra from stars that span the HR diagram (\teff\ = 3500--7500 K;\logg\ = 2.0--5.0). Parameters for the library stars are determined from LTE spectral modeling. Once the target spectrum and library spectrum are placed on the same wavelength scale, SpecMatch computes the sum of the squares of the pixel-by-pixel differences in normalized intensity. The weighted mean of the ten spectra with the lowest values is taken as the final value for the effective temperature, stellar surface gravity, and metallicity. SpecMatch-derived stellar radii are uncertain to 10\% RMS, based on tests of stars having known radii from high resolution spectroscopy and asteroseismology \citep{huber2013}. The stellar parameters determined using SpecMatch were \teff\ = $5757\pm60$\,K, \logg\ = $4.32\pm0.07$\,dex and  \feh\ = $0.21\pm0.04$.

Two other analyses were carried out on the HIRES spectrum to better understand the role of instrumental systematic errors between the three sets of observations, and to better determine how to set the error bars on the resulting stellar parameters.

An SPC analysis of the HIRES spectrum gave the following stellar parameters: \teff\ = $5740\pm50$\,K, \logg\ = $4.42\pm0.10$\,dex, \feh\ = $0.22\pm0.08$, \vsini\ = $0.8\pm0.5$\,\kms, maximum height of the CCF = 0.988, and S/N = 58.8 per resolution element.

We also  performed a standard EW analysis of FeI and FeII lines using MOOG and Kurucz's odfnew models. We determined two sets of \teff, \logg, \feh, and \vt\ (microturbulence) values, one using only the Keck spectrum and a second using a solar spectrum from MIKE/Magellan for line-by-line differential analysis. The solution based solely on the Keck spectrum gave \teff\ = $5740\pm35$\,K, \logg\  = $4.230\pm0.072$\,dex, A(Fe) = $7.649\pm0.033$ (which would correspond to \feh\ = +0.2 for an absolute solar iron abundance of A(Fe) = 7.45, as inferred by numerous 1D-LTE analyses), and \vt\ = $1.03\pm0.08$\,\kms. The solution with  the Keck spectrum + MIKE/Magellan solar spectrum template gave \teff\ = $5818\pm21$\,K, \logg\ = $4.33\pm0.05$\,dex, \feh\ = $0.24\pm0.02$, and micro turbulent velocities, \vt\ = $1.07\pm0.05$\,\kms. 

As an additional search for unseen companions, we searched the HIRES spectrum for secondary spectral lines caused by a second star that is angularly close enough to the primary to fall into the HIRES spectrometer slit (0\farcs87 wide). The observed spectrum is first compared to a library of representative main sequence stars. The best match library spectrum is subtracted from the observed spectrum, and the residuals are then inspected for evidence of a secondary spectrum \citep{kolbl2015}. We found no secondary spectrum brighter than 1\% of the primary, separated in radial velocity by more than 10\,\kms. Stars separated by less than 10\,\kms\ cannot be detected because they overlap with the primary spectrum. See section~\ref{s:blender} for more details.

Table~\ref{table:spectroscopy} contains the results for the spectroscopy of \koicurCCkic\ and attendant analyses. We considered the scatter and precision in the spectroscopic results in order to determine the most reasonable stellar parameters for the recovery of planetary parameters, as described in the next section.

\begin{deluxetable}{lcc}
\tabletypesize{\scriptsize}
\tablewidth{0pc}
\tablecaption{Spectroscopy and Analysis Results for \koicurCCkic\  \label{table:spectroscopy}}
\tablehead{\colhead{Parameter}	& 
\colhead{Value} 		& 
\colhead{Notes}}
\startdata
\sidehead{\em Tull/McDonald Observatory}
Effective temperature (K) 				& 5650 $\pm$ 108 	& 	a\\
Surface gravity \logg\ (cgs)				& 4.45 $\pm$ 0.16	& 	a\\
Metallicity \feh (dex)					& 0.21 $\pm$ 0.07 	& 	a\\
Projected rotation Velocity \vsini\ (\kms)	& 4.3 $\pm$ 0.5		& 	a\\

\sidehead{\em TRES/Whipple Observatory}
Effective temperature (K)	 			& 5751 $\pm$ 55		& 	b\\
Surface gravity \logg\ (cgs)				& 4.43 $\pm$ 0.10	& 	b\\
Metallicity \feh (dex)					& 0.40 $\pm$ 0.08 	& 	b\\
Projected rotation Velocity \vsini\ (\kms)	& 4.2 $\pm$ 0.5		& 	b\\

\sidehead{\em HIRES/Keck Observatory}
Effective temperature (K)     			& 5757 $\pm$ 60		& 	c\\
Surface gravity \logg\ (cgs)				& 4.32 $\pm$ 0.07	& 	c\\
Metallicity \feh (dex)					& 0.21 $\pm$ 0.04	& 	c\\
Projected rotation Velocity \vsini\ (\kms)	& $<$1				& 	c \\
\vspace{-.1cm}\\
Effective temperature (K)	  			& 5740 $\pm$ 50		& 	b\\
Surface gravity \logg\ (cgs)				& 4.42 $\pm$ 0.10	& 	b\\
Metallicity \feh (dex)					& 0.22 $\pm$ 0.08	& 	b\\
Projected rotation Velocity \vsini\ (\kms)	& 0.8 $\pm$ 0.5		& 	b \\
\vspace{-.1cm}\\
Effective temperature (K)     			& 5818 $\pm$ 21		& 	d\\
Surface gravity \logg\ (cgs)				& 4.33 $\pm$ 0.05	& 	d\\
Metallicity \feh (dex)					& 0.24 $\pm$ 0.02	& 	d\\

\enddata
\tablecomments{\\
a: Analyzed with \kea.\\
b: Analyzed with SPC.\\
c: Analyzed with SpecMatch.\\
d: Analyzed with MOOG.\\
}
\end{deluxetable}

\section{Host Star Properties}\label{s:hostStarProperties}

\begin{figure}
\resizebox{\hsize}{6.5in}{\includegraphics{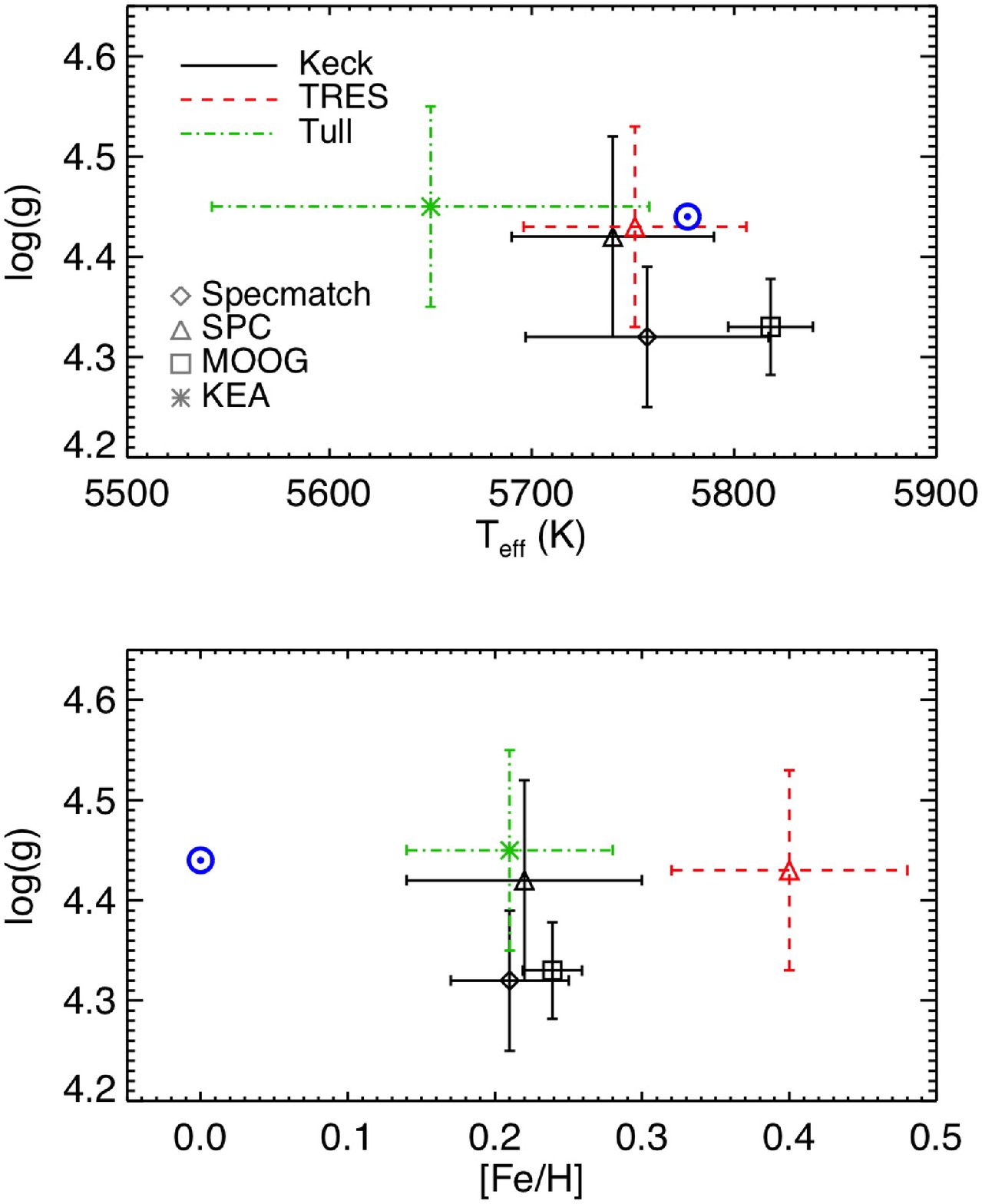}}
\caption{Spectroscopically derived surface gravity as a function of effective temperature (top panel) and metallicity (bottom panel) for \koicurCCkic, derived using different methods and spectra (see text for details). The position of the Sun is also shown.}
\label{fig:specresults}
\end{figure}

Since \koicurCCkic\ is too faint for direct observations such as interferometry or 
asteroseismology, the derivation of stellar mass, radius and density relies on matching atmospheric properties to interior models. Figure \ref{fig:specresults} compares the spectroscopic solutions described in the previous section. The values are generally within 1\,$\sigma$ and no strong correlation of \logg\ with \teff\ or \feh\ is 
apparent, indicating that the differences are dominated by differences in the methods rather than degeneracies in atmospheric properties \citep[e.g.,][]{torres2012}. On average, \koicurCCkic\ is slightly cooler, slightly larger, and about 60\% more metal-rich than the Sun. 

To ensure self-consistency, we adopted a single spectroscopic solution for the remainder of the paper. We chose the SpecMatch solution, which yields an intermediate temperature, conservative surface gravity, and metallicity consistent with most other methods. To account for the systematic differences between methods, we added the standard deviation of all methods for a given parameter in quadrature to the formal errors provided by SpecMatch. The final adopted atmospheric properties are listed in Table \ref{tab:stprop}.

To derive interior properties, we fitted \teff, \logg, and \feh\ to a grid 
of Dartmouth isochrones \citep{Dotter08}. The grid was calculated using 
solar-scaled alpha-element abundances, and interpolated to a stepsize of 0.01\,\msun\ in mass and 0.02\,dex metallicity. We fitted the input values to the model grid using a Markov chain Monte Carlo (MCMC) algorithm which, at each step, draws samples in mass, age, and metallicity. For ease of computation, the samples in age and 
metallicity were drawn in discrete steps corresponding to the sampling of the model grid (0.5\,Gyr in age and 0.02\,dex in \feh). For each sample of fixed age and metallicity, we linearly interpolated the grid in mass to derive \teff\ and \logg, which were then used to evaluate the likelihood functions. We adopted uniform priors in mass, age, and metallicity. We calculated $10^6$ iterations and discarded the first 10\% as burn-in.

\begin{figure*}
\resizebox{\hsize}{!}{\includegraphics{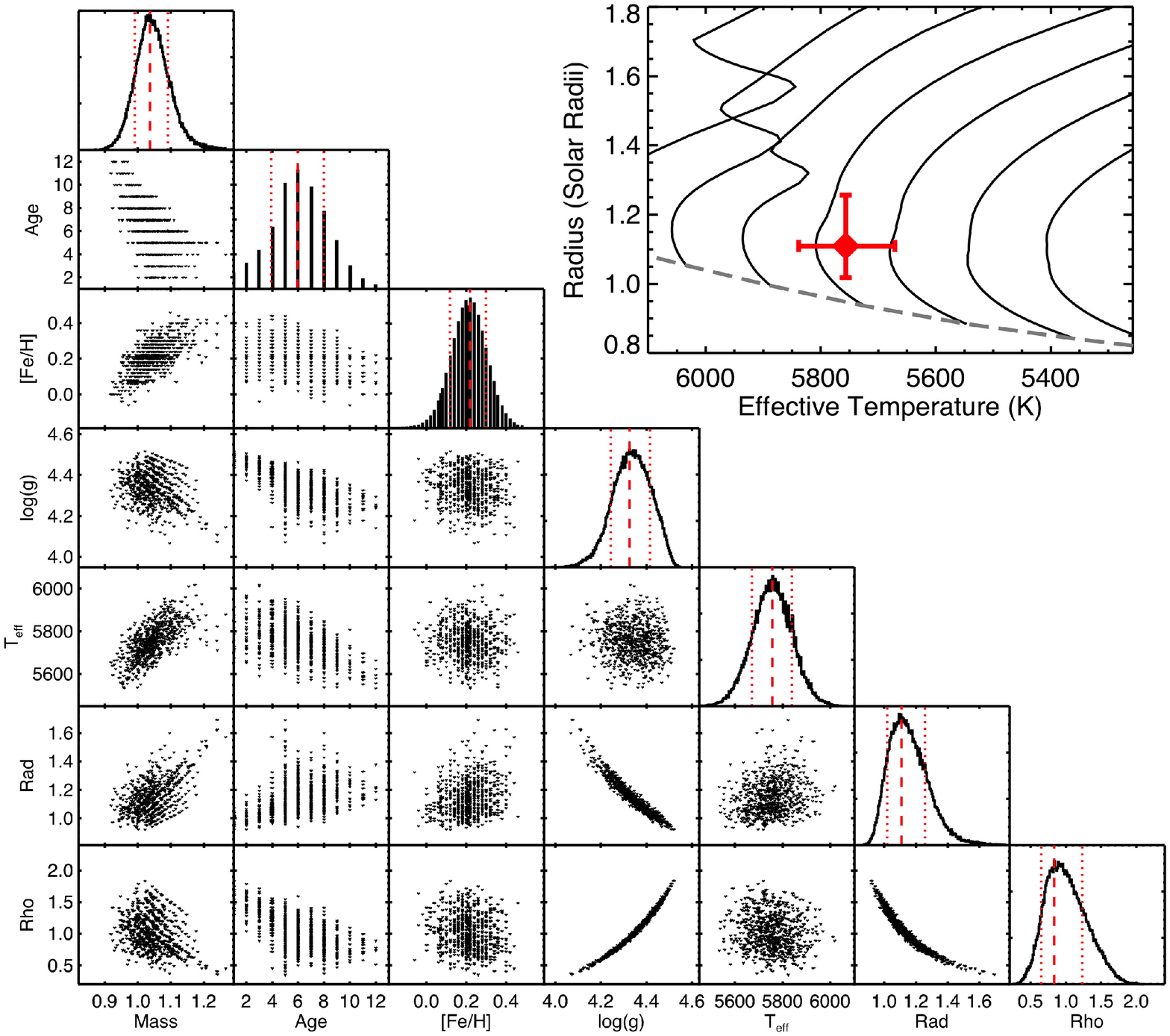}}
\caption{Marginalized and full posterior distributions of Markov chain Monte Carlo chains based on using spectroscopic \teff, \logg\ and \feh\ to constrain a grid of Dartmouth isochrones. A random set of 750 samples ($\sim 0.1$\% of all iterations) is plotted for each parameter for clarity. Vertical dashed and dotted lines mark the 
posterior mode and 68\% confidence interval around the mode for each parameter. 
Note that age and metallicity are discretely sampled with the resolution of the 
model grid. The top right panel shows the position of \koicurCCkic\ in a radius temperature diagram, with solid black lines showing evolutionary tracks with 0.9 -- 1.2 \msun\ in steps of 0.05 \msun\ for \feh\ = 0.21. The gray dashed line marks the zero-age main-sequence.}
\label{fig:stprop_mcmc}
\end{figure*}

Figure \ref{fig:stprop_mcmc} shows the marginalized and full posterior distributions, as well as the position of \koicurCCkic\ on a radius-temperature diagram. As expected, \koicurCCkic\ is slightly evolved off the zero-age main-sequence. The 68\% confidence interval for the stellar radius ranges from 1.02--1.26\,\rsun, with a most probable value of 1.11\,\rsun.

\begin{deluxetable}{lc}
\tabletypesize{\scriptsize}
\tablewidth{0pc}
\tablecaption{Stellar Properties of \koicurCCkic \label{tab:stprop}}
\tablehead{\colhead{Parameter}	& 
\colhead{Value} 		
}
\startdata
\sidehead{\em Derived from high-resolution spectroscopy using SpecMatch}
Effective Temperature (K)		& \KOIteff\		\\
log [Surface gravity] (dex)		& \KOIlogg\		\\
Metallicity \feh\ (dex)			& \KOIfeh\		\\

\sidehead{\em Derived from fitting \teff, \logg\ and \feh\ to isochrones}
Radius (\rsun)					& \KOIrstar		\\
Mass (\msun)					& \KOImstar	\\
Mean Density (g\,cm$^{-3}$)		& \KOIrhostar	\\
Age (Gyr)						& $\sim6\pm2$	\\
\enddata
\end{deluxetable}

As a consistency check, we also computed the age, mass, and radius of this star based on the MOOG analysis of the HIRES spectrum using Yonsei-Yale isochrones \citep{Demarque04,Yi2004} and the approach detailed in Section 4.5 of \citet{ramirez2014}, obtaining an  age of 5.2 (3.4--7.5) Gyr, a mass of 1.07 (1.02--1.12) \msun, and a radius of 1.12 (1.03--1.29) \rsun. The values in parentheses represent approximate $\pm 1\, \sigma$ ranges of values. 
Note that the agreement in these derived quantities with those derived for the SpecMatch results is very high, giving us confidence that we have a reasonable degree of control over the systematics in the different analysis techniques.

\section{Properties of the Planet \koicurb}\label{s:planet properties}

To estimate the posterior distribution on each fitted parameter in our transit model, we used an MCMC approach similar to the procedure outlined in \citet{ford2005} and implemented in \citet{roweMultis2014}.   Our algorithm used both a Gibbs sampler and a control set of parameters to enable a vector jumping distribution to efficiently handle the sampling of correlated parameters as outlined in \citet{gregory2011}.  We directly fitted the following transit and orbital parameters: orbital period ($P$), epoch, scaled planet radius ($\slfrac{\rpl}{\rstar}$), impact parameter ($b$), and eccentricity parameters (\ecosw\ and \esinw). We generated ten sets of 1,000,000-element Markov chains, then discarded the first 20\% of each chain to establish burn-in. The remaining sets were combined and used to calculate the mode, standard deviation, and $1\,\sigma$ bounds of the distribution centered on the mode of each model parameter.  Our model fits and uncertainties are reported in Table~\ref{table:parameters}.   The eccentricity of \koicurb\ was not well constrained by the light curve as was expected in the absence of a radial velocity detection of the planet.  

We used the Markov chains to derive model dependent measurements of the transit depth, \Tdep, and transit duration, \Tdur.  The transit depth is defined as the transit-model value when the projected distance between the star and planet is minimized.  The transit duration is the time from first to last contact.  
We also convolved the transit model parameters with the stellar parameters (see Section~\ref{s:hostStarProperties}) to compute the planetary radius, \rpl\ = \KOIRpshort\ \rearth, and the flux received by the planet relative to the Earth ($S=\KOIinsolationshort\ \Searth$). We also performed a fit forcing the orbit to be circular in our transit model and obtained thereby an estimate for the density of the star of \rhostar\ = $1.1\pm0.3$\,\gcmc. While this value is perfectly consistent with the spectroscopic work, it suggests that both the star and the planet may be somewhat smaller than indicated by the SpecMatch analysis.  

\begin{deluxetable}{lcc}
\tabletypesize{\scriptsize}
\tablewidth{0pc}
\tablecaption{Planet Parameters for \koicurb \label{table:parameters}}
\tablehead{\colhead{Parameter}	& 
\colhead{Value} 		& 
\colhead{Notes}}
\startdata

\sidehead{Transit and orbital parameters}
Orbital period $P$ (day)						& \KOIperiod		& a, b	\\
Epoch (BJD - 2454833)						& \KOIepoch		& a, b \\
Scaled planet radius $\slfrac{\rpl}{\rstar}$					& \KOIrpRstar		& a, b	\\
Impact parameter $b \equiv \slfrac{a \cos{i}}{\rstar}$		& \KOIimpactParameter		& a, b	\\
Orbital inclination $i$ (deg)					& \KOIinclination 		& a	\\
Transit depth \Tdep\ (ppm)					& \KOIdepth & a\\
Transit duration \Tdur\ (hr)						& \KOIduration & a\\
Eccentricity \ecosw\							& \KOIecosw & a, b\\
Eccentricity \esinw\							& \KOIesinw & a, b\\

\sidehead{Planetary parameters}
Radius \rpl\ (\rearth)							& \KOIRp		& a	\\
Orbital semimajor axis $a$ (AU)				& \KOIa		& a 	\\
Equilibrium temperature \teq\ (K)				& \koiTequ		& c \\
Insolation relative to Earth						& \KOIinsolation    	& d\\
\enddata

\tablecomments{\\
a: Based on the photometry.\\
b: Directly fitted parameter. \\
c: Assumes Bond albedo = 0.3 and complete redistribution.\\
d: Based on Dartmouth isochrones.}
\end{deluxetable}

\section{Adaptive Optics Observations}\label{s:adaptiveoptics}

\koicurCCkic\ has one companion visible in the UKIRT image (see Figure~\ref{fig:diffImages}) at a separation of $\sim$4\arcsec. This dim \Kp\ = 18.435 star is well outside the 3 $\sigma$ confusion radius and thus cannot be the source of the transit-like features in the \kepler\ photometry, according to the centroid arguments presented in Section~\ref{ss:centroidanalysis}. However, it is important to probe the scene around \koicurCCkic\ for evidence of nearby companions that would not be identifiable at the spatial resolutions in either the \kepler\ or the UKIRT image data.

We obtained near-infrared adaptive optics (AO) images in the $J$ (1.248 $\mu$m) and Br-$\gamma$ (2.157 $\mu$m) filters using the NIRC2 imager behind the natural guide star AO system \citep{wizinowich2004,johannson2008} at the Keck II telescope on UT 2014, June 13.  Data were acquired in a three-point dither pattern to avoid the lower left corner of the NIRC2 array, which is noisier than the other three quadrants.  The dither pattern was performed three times with step sizes of 3\farcs0, 2\farcs5, and 2\farcs0 for a total of nine frames; individual frames had an integration time of 60~s (30~s times 2 coadds) in both bands.  The individual frames were flat-fielded, sky-subtracted (sky frames were constructed from a median average of the 9 individual source frames), and co-added.

\begin{figure*}[ht]
\resizebox{\hsize}{!}{\includegraphics{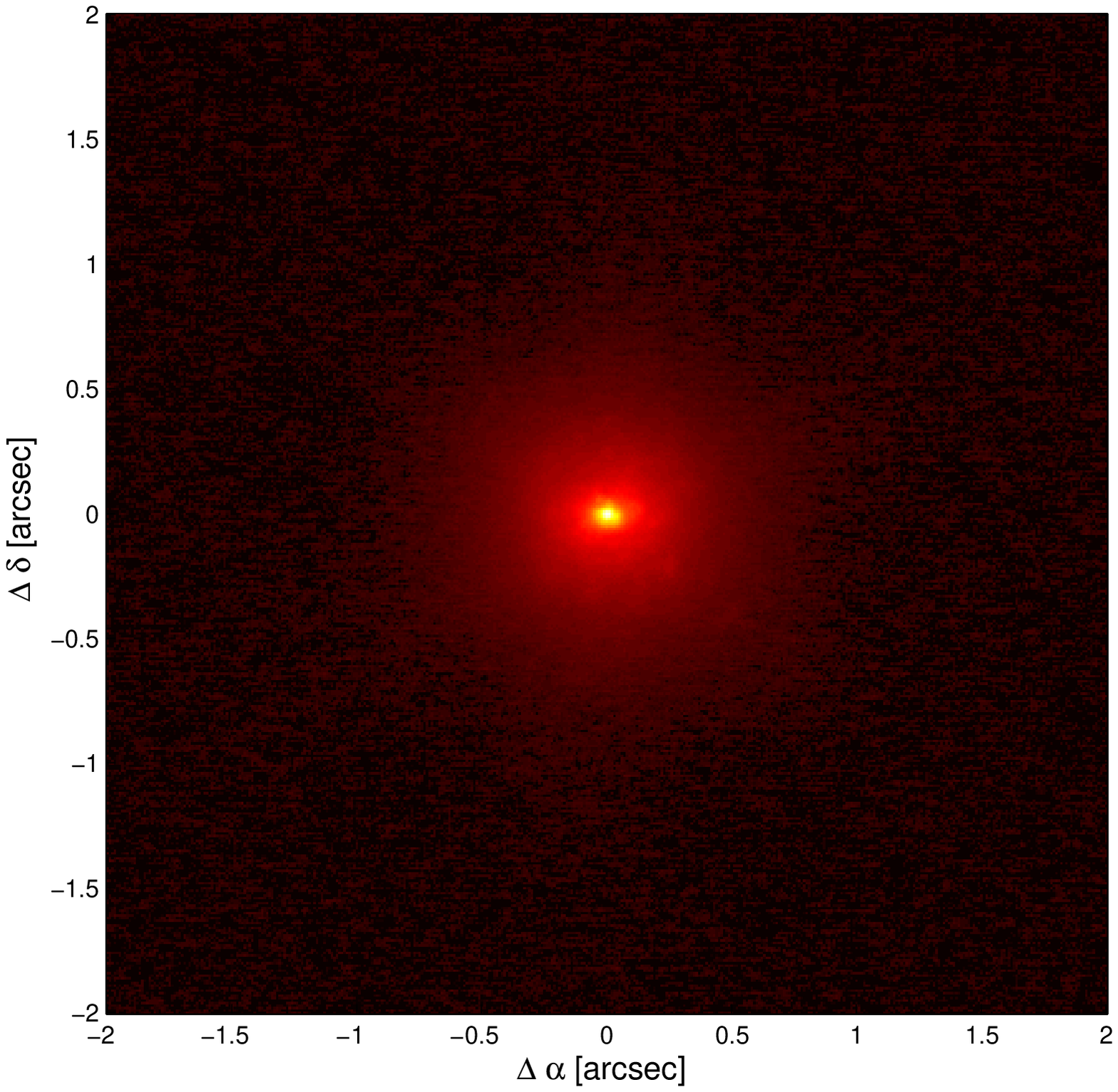}
\includegraphics{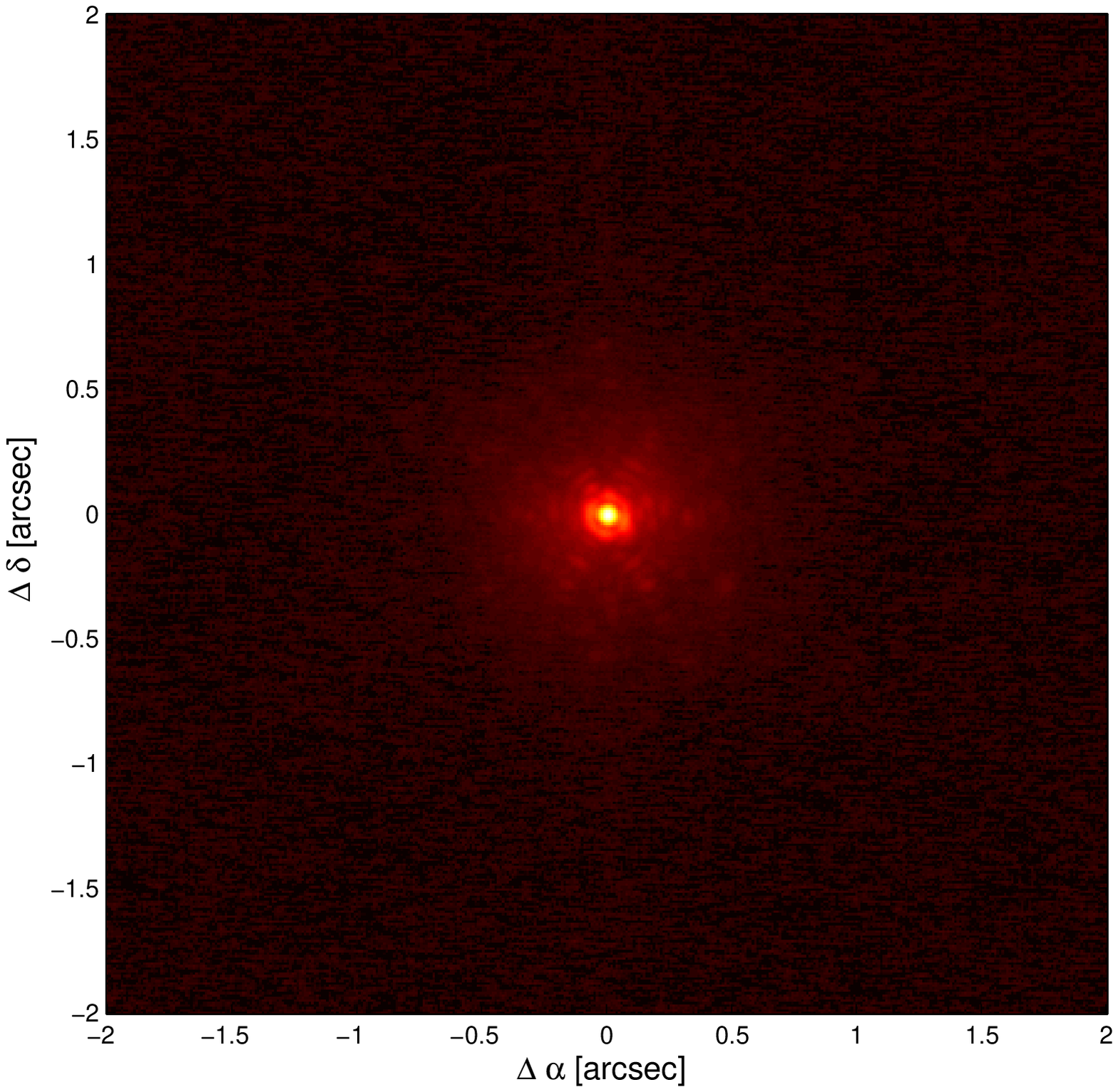}}\\
\caption{
Images obtained with the NIRC instrument at the Keck II telescope on 2014 June 13 in $J$ band (left) and $K$ band (right). Neither band indicates the presence of any background stars that might be the source of the transits observed in the \kepler\ photometry of \koicur. 
\label{fig:keckaoimages}}
\end{figure*}

\begin{figure*}[ht]
\resizebox{\hsize}{!}{
\includegraphics{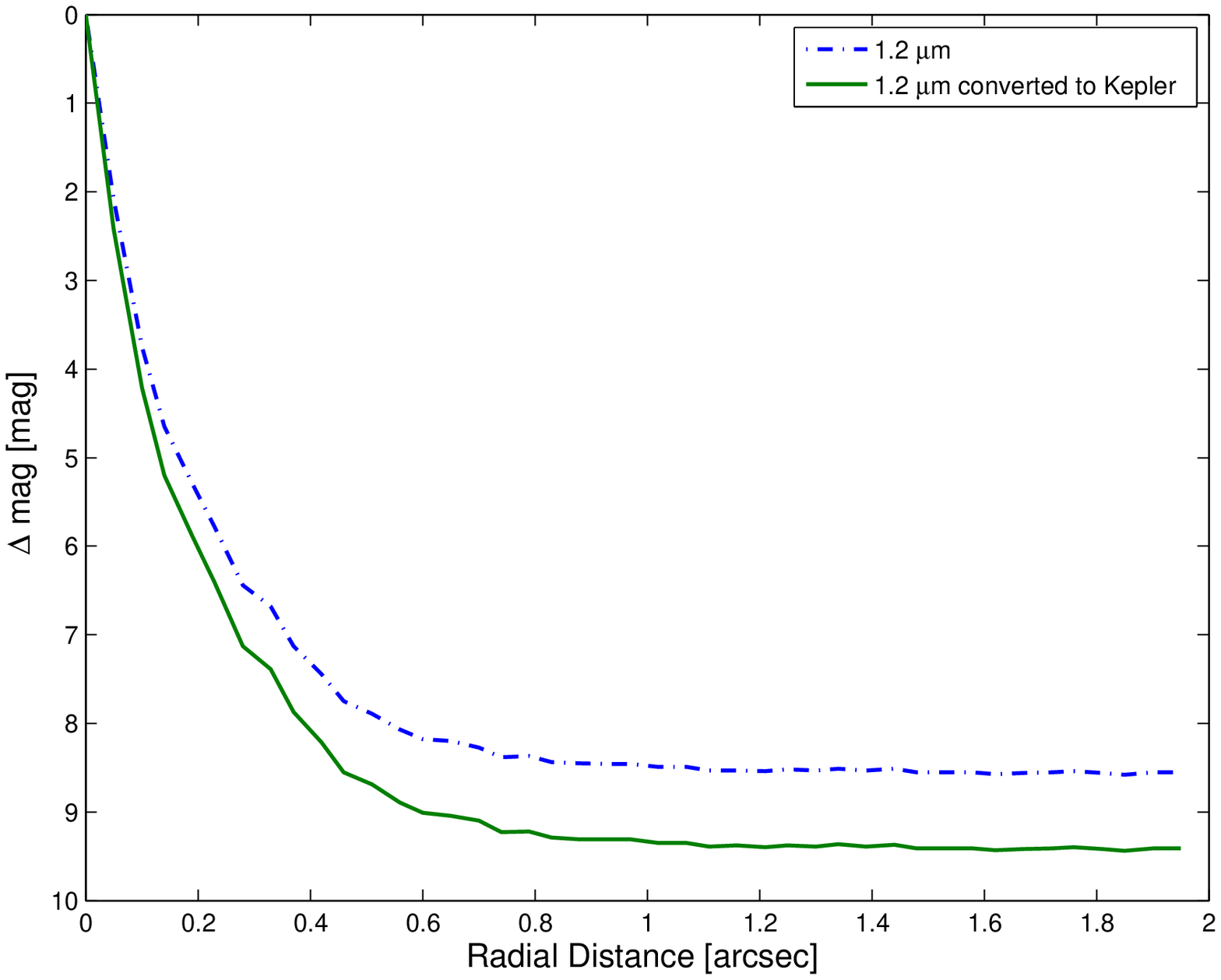}
\hspace{.3cm}
\includegraphics{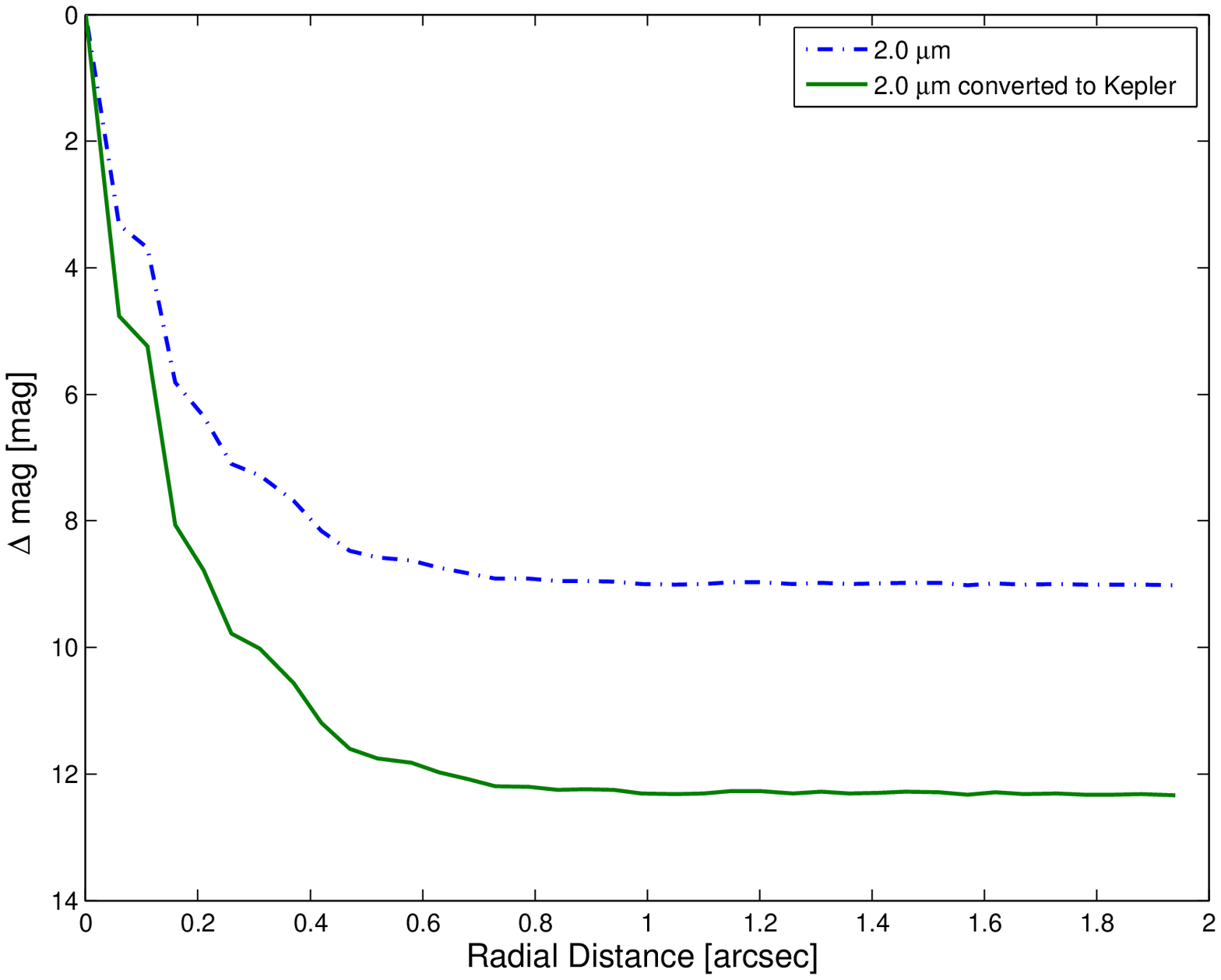}}\\
\caption{
5\,$\sigma$ sensitivity limits on the rejection of background confusion sources in terms of $\Delta\rm{mag}$ as a function of radial offset from \koicurCCkic. Left: the dashed-dotted blue curve is the measured in-band sensitivities at $1.2\,\mu$m, while the solid green curve indicates the corresponding estimated sensitivity limits converted to the \kepler\ bandpass. Right: same sensitivity curves for $2.0\,\mu$m
\label{fig:keckaolimits}}
\end{figure*}

The average AO-corrected seeing (FWHM) was 0\farcs04635 and 0\farcs05229 at $J$- and $K$-bands, respectively. Figure~\ref{fig:keckaoimages} shows the Keck AO images obtained for this star. \koicurCCkic\ appears to be a well isolated star in both bands with no obvious background stars, or resolved, associated companions within $\pm$2\arcsec\ that would present blends at the 4\arcsec\ pixel scale of \kepler's camera. Figure~\ref{fig:keckaolimits} shows the 5\,$\sigma$ sensitivity limits for rejecting the presence of background confusion sources as a function of radial distance from \koicurCCkic. The sensitivity achieved in both bands establishes rejection of background stellar sources from $\sim$0\farcs5 to 2\farcs0 within $\Delta\rm{mag} \approx 9$\,mag of the central target star, in the bands of observation.  Using typical stellar \kepler-infrared colors \citep{howell2011}, the limitations of the observations translate into the \kepler\ bandpass sensitivity limits near $\Delta\rm{mag} \approx$10--12\,mag. Given that the transits of \koicurb\ are \KOIdepthshort\,ppm deep, a background eclipsing binary exhibiting 50\% eclipses would need to be within 8.66 mag of \koicur. The AO images and the difference image analysis carried out with the \kepler\ data allow us to validate the planetary nature of the transit source with high confidence, as detailed in the next section.

\section{\blender\ analysis}
\label{s:blender}

In this section, we examine the possibility that the \koicurb\ signal may be due to an astrophysical false positive. Typical examples of such phenomena include a background or foreground eclipsing binary along the line of sight to the target (i.e., within the photometric aperture of \kepler), a chance alignment with a background or foreground single star transited by a (larger) planet, or a physically associated star transited by a planetary or stellar companion (hierarchical triple). We refer to the first and second of these scenarios as ``BEB'' and ``BP,'' respectively, and to the third as ``HTP'' or ``HTS,'' depending on whether the object producing the eclipses is a planet or a star.\footnote{Note that in this paper we consider the BP and HTP scenarios as false positives, even though they involve  a transiting planet. This is because, for a given transit signal, a planet orbiting an unseen star in the aperture could be considerably larger than if it orbited the target, and therefore less interesting for our purposes, as the likelihood of its being rocky would be much smaller.}

Other possible false positive scenarios include the transit of a white dwarf, or the occultation of a brown dwarf in an eccentric orbit with an orientation such that the transit does not occur \citep[see, e.g.,][]{Santerne:13}. The white dwarf scenario can be ruled out because it would result in a very large microlensing signal of $\sim$2400\,ppm, as per \citet{sahuAndGilliland2003}, that is not observed. A non-transiting, occulting white dwarf is unlikely to produce a 200-ppm deep transit-like feature as the white dwarf would need to be about 7000 K to produce an occultation of the correct depth, and white dwarfs cool from 10,000 K to 6000 K in $\sim$2 Gyr \citep[see, e.g.,][]{goldsbury2012}. Thus, there would be only a very brief opportunity for a white dwarf to provide 200-ppm deep occultations of \koicurCCkic\ as the uncertainty in the transit depth is $\sim$10\%. Further, we expect white dwarf/main sequence pairs at a period of 385 days to be relatively rare: \citet{farmerAndAgol2003} predicted that only five microlensing white dwarf/main sequence binaries would be discovered in \kepler's FOV out to periods of $\sim$100 days, and indeed, only one such white dwarf has been found to date \citep{kruseAndAgol2014} in an 88 day period orbit of a solar-like star exhibiting $\sim$1000\,ppm microlensing events and (coincidentally) $\sim$1000\,ppm deep secondary occupations. 

The brown dwarf scenario is very unlikely to begin with (much more so than the BEB, BP, or HTP/HTS scenarios) because of the known paucity of brown dwarfs as spectroscopic companions to main-sequence stars within hundreds of AU, and especially G-type stars \citep[see, e.g.,][]{marcyAndButler2000}, whereas binaries with main-sequence companions are far more common. Only eight transiting brown dwarf/main sequence binaries have been discovered to date \citep{moutou2013}, four of which are in the \kepler\ FOV \citep{bouchy2011,johnson2011,diaz2013,moutou2013}. All but one of these have orbital periods under 20 days, and KOI-415b has a period of 167 days \citep{moutou2013}. Further, even the most massive brown dwarf would cool below a luminosity ratio of 100\,ppm with the Sun in 1.5 Gyr , and below 25\,ppm by 6 Gyr \citep{burrows2001}. Even if such a case were to occur, the shape of the transit-like signals (particularly the ingress/egress) would not match what we observe because of the much larger size of brown dwarfs compared to terrestrial planets like Kepler-NNNb. Thus we restrict our analysis to the above BEB, BP, and HTP/HTS configurations.

The general procedure we followed consists of generating realistic transit-like light curves for a large number of these false positive scenarios (``blends''), and comparing each of them with the \kepler\ photometry for \koicur\ to see which provide a fit indistinguishable from that of a model of a planet transiting the target star. This comparison allowed us to place constraints on the various physical parameters of the blends, including the sizes and masses of the bodies involved, their overall color and brightness, the linear distance between the background/foreground eclipsing pair and the target, etc.  We then used a Monte Carlo approach to estimate the expected rate of occurrence of each type of blend (BEB, BP, HTP/HTS) from our simulations, and we compared the sum with the expected rate of occurrence of transiting planets similar to the one implied by the \koicur\ signal.  If the \emph{a priori} planet occurrence is much larger than the anticipated total blend frequency, then we would consider the candidate to be statistically validated.

The algorithm used to generate blend light curves and compare them to the observed photometry in a $\chi^2$ sense is known as \blender\ \citep{torres2004, torres2011, fressin2012}, which has been applied in the past to the validation of many of the most interesting \kepler\ discoveries \citep[see, e.g.,][]{Ballard13, Barclay13, kepler62,  Meibom13, Kippingetal14b}, including, most recently, a sample of twelve small planets in the habitable zone of their parent stars \citep{torres2015}. For full details of the technique, we refer the reader to the latter publication. That work also explains the calculation of the blend frequencies through a Monte Carlo procedure, making use of the constraints provided by \blender\ along with other constraints available from follow-up observations. These include the sensitivity to companions at close angular separations from our AO imaging (Section~\ref{s:adaptiveoptics}), limits on the brightness of unseen companions at even closer separations derived from our high-resolution spectroscopy (Section~\ref{ss:HIRES}), constraints on blended objects from our difference image analysis (specifically, the size of the angular $3\,\sigma$ exclusion region;
Section~\ref{ss:centroidanalysis}), and also the brightness measurements for \koicurCCkic\ as listed in the KIC (which allowed us to rule out blends whose combined light is either too red or too blue compared to the observed color of the target).

The time series used for the \blender\ analysis is the detrended Q1--Q17 LC systematic error-corrected light curve for \koicurCCkic. The stellar parameters adopted for the target were those given in Section~\ref{s:hostStarProperties}. The parameters of the simulated stars in the various false positive configurations were drawn from model isochrones from the Dartmouth series \citep{Dotter08}, and rely also on the known distributions of binary properties such as the orbital periods, mass ratios, and eccentricities \citep{Raghavan10, Tokovinin14}.  Other ingredients used for the blend frequency calculation included the number density of stars in the vicinity of the target from the Besan\c{c}on Galactic structure model \citep{Robin:03,  Robin12},\footnote{\url{http://model.obs-besancon.fr}\,.} and the rates of occurrence of eclipsing binaries \citep{Slawson11} and of transiting planets from the \kepler\ Mission itself (NASA Exoplanet Archive, consulted on 2014 March 16).\footnote{\url{http://exoplanetarchive.ipac.caltech.edu}\,.}

\begin{figure}[ht]
\resizebox{\hsize}{!}{\includegraphics{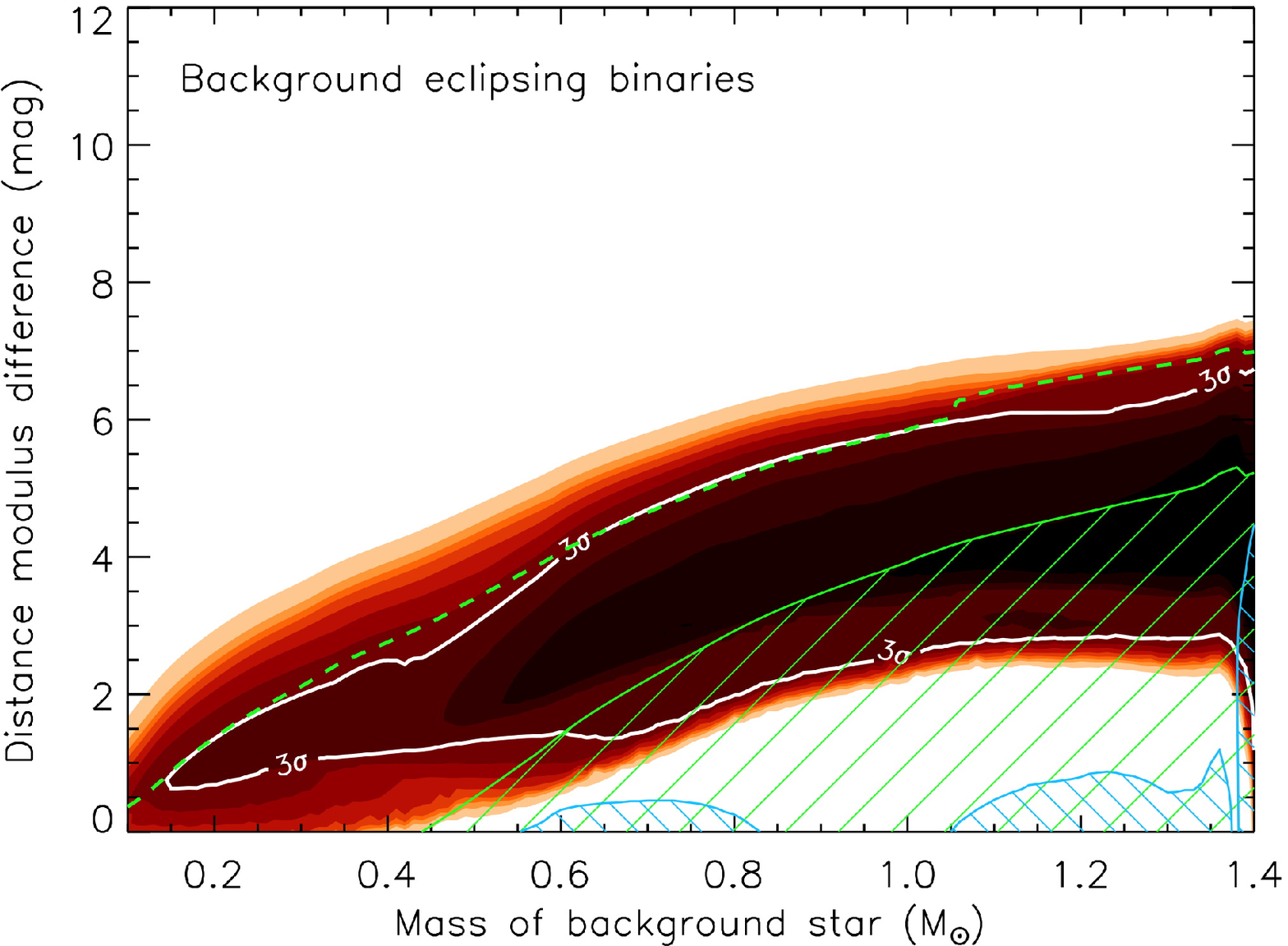}}
\caption{
Map of the $\chi^2$ surface (goodness of fit) for \koicurCCkic\ corresponding to blends involving background eclipsing binaries. On the vertical axis we represent the linear distance between the BEB and the target $\left(D_{\rm BEB}-D_{\rm \rm targ}\right)$, cast for convenience in terms of the difference in distance modulus, $\Delta\delta = 5 \log\left(D_{\rm BEB}/D_{\rm targ}\right)$. Only blends within the solid white contour (darker colors) provide fits to the \kepler\ light curve that are within acceptable limits \citep[$3\,\sigma$, where $\sigma$ is the significance level of the $\chi^2$ difference compared to a transiting planet model fit; see ][]{fressin2012}. Other concentric colored areas (lighter colors) represent fits that are increasingly worse ($4\,\sigma$, $5\,\sigma$, etc.), which we considered to be ruled out. The blue cross-hatched areas correspond to regions of parameter space where the blends are either too red (left) or too blue (right) compared to the measured $r - K_s$ color of the target, by more than three times the measurement uncertainty. The dashed green line labeled $\Delta \Kp$ = 8.4 is tangent to the white contour from above, and corresponds to the faintest viable blends. The green line labeled $\Delta \Kp$ = 5.0 represents the spectroscopic limit on faint background stars. All simulated blends below this line (green hatched region) are brighter and are generally excluded if the BEB is angularly close enough to the target to fall within the slit of the spectrograph.
\label{fig:blenderBEB}}
\end{figure}

\begin{figure}[ht]
\resizebox{\hsize}{!}{\includegraphics{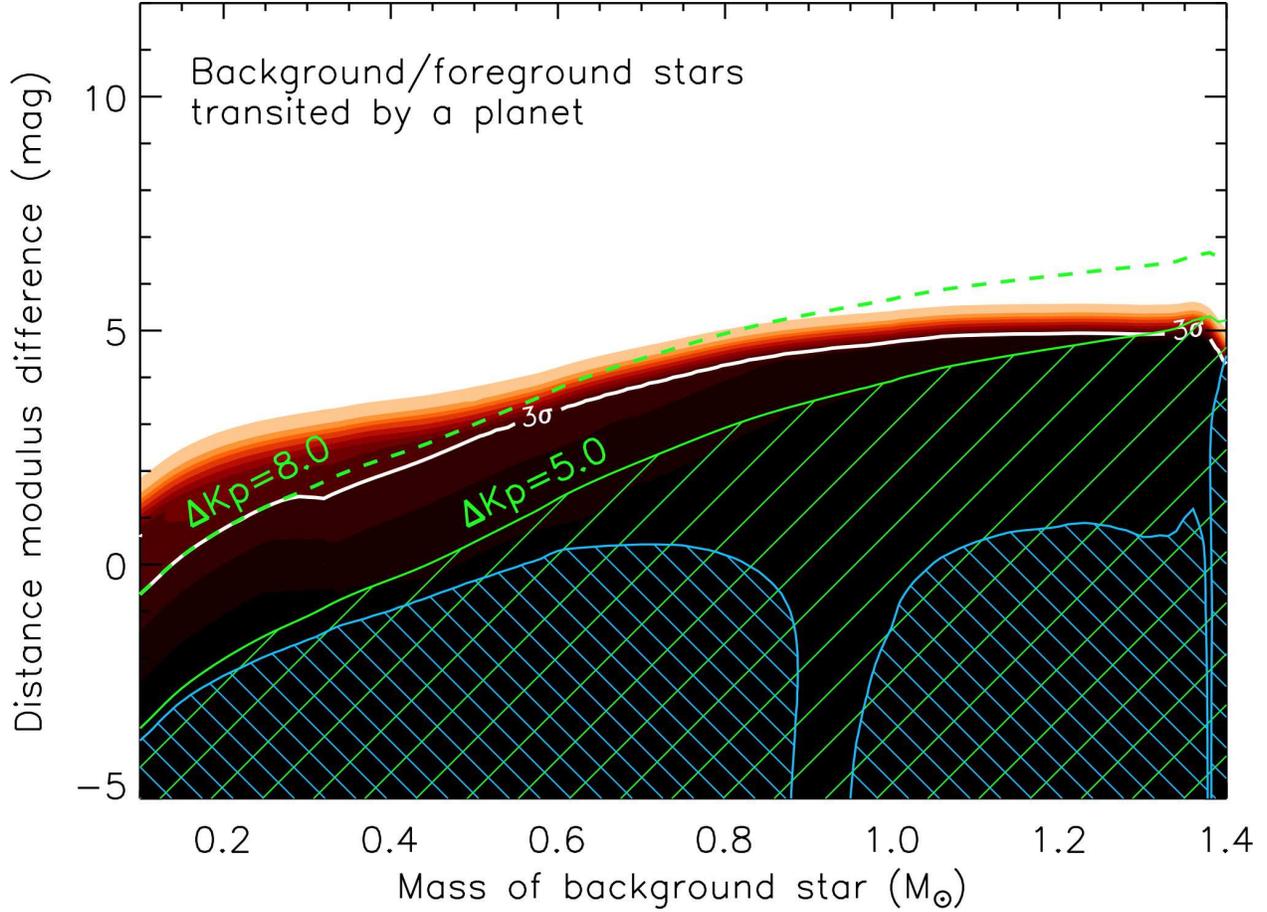}}
\caption{
Similar to Figure~\ref{fig:blenderBEB} (and with the same color scheme) for blends involving background or foreground stars transited by a planet (BP scenario). The faintest blends giving acceptable fits have $\Delta \Kp$ = 8.0 relative to the target (dashed green line). Blends below the solid green ($\Delta \Kp$ = 5.0, hatched green area) are generally excluded by the spectroscopic constraint.
\label{fig:blenderBP}}
\end{figure}

\begin{figure}[ht]
\resizebox{\hsize}{!}{\includegraphics{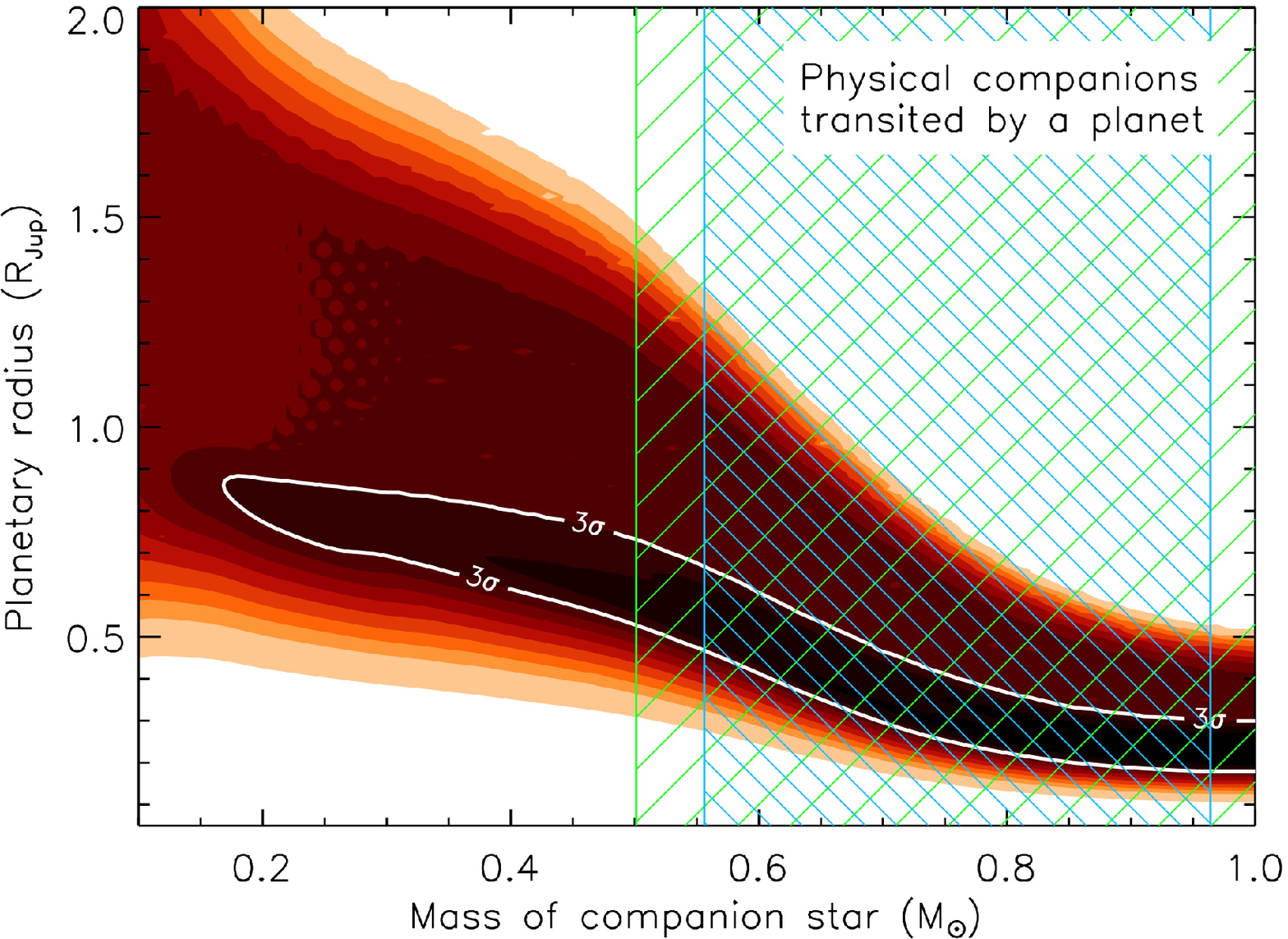}}
\caption{
Similar to Figure~\ref{fig:blenderBEB} for the case of physical companions to \koicurCCkic\ that are transited by a planet (HTP). Only companion stars with masses larger than about 0.96\,$\msun$ or smaller than about 0.56\,$\msun$ yield blend colors that are consistent with the measured $r - K_s$ index of \koicurCCkic. However, some of these blends are eliminated by the spectroscopic constraint, as they are brighter than $\Delta \Kp$ = 5.0 (green hatched area). The only false positives that would remain undetected (viable) have companion stars smaller than 0.50\,$M_{\sun}$ and are transited by planets with sizes that place them within the white contours.
\label{fig:blenderHTP}}
\end{figure}

The \blender\ constraints for the BEB, BP, and HTP scenarios are illustrated in Figures~\ref{fig:blenderBEB}--\ref{fig:blenderHTP}. Our simulations indicate that eclipsing binaries orbiting the target (HTS scenario) invariably produce light curves that have the wrong shape for a transit, or feature noticeable secondary eclipses that are not seen in the \koicurCCkic\ photometry.  Similarly, blends involving background giant stars yield light curves that also have the wrong
shape. We therefore restricted our analysis to main-sequence stars. Figure~\ref{fig:blenderBEB} displays a sample cross-section of the blend parameter space for the BEB scenario, with the horizontal axis showing the mass of the primary star in the background eclipsing binary, and the vertical axis representing the relative distance between the background binary and the target.  Blends with parameters interior to the white contour give a goodness of fit to the \kepler\ photometry that we considered indistinguishable from that of a transiting planet fit (to within $3\,\sigma$). We consider these as viable false positives. Also indicated in the figure are regions in which blends have an $r-K_s$ color index that does not match that of the target (see figure caption), or where blends are not allowed because the intruding EB would have been detected spectroscopically. Similar diagrams for the BP and HTP configurations are shown in Figure~\ref{fig:blenderBP} and Figure~\ref{fig:blenderHTP}. Further explanations regarding the \blender\ constraints may be found in the recent work by \cite{torres2015}.

Based on the above analysis, and including also the information derived from the follow-up observations mentioned earlier, we arrived at estimates of the blend frequencies for the BEB, BP, and HTP scenarios of $1.21 \times 10^{-12}$, $2.56 \times 10^{-10}$, and $2.35 \times 10^{-6}$, respectively. These estimates make use of the frequencies of eclipsing binaries generated by the \kepler\ team \citep{Slawson11} as well as the frequencies of transiting planets of the specific size required to produce the above blends, based on the available sample of KOIs from the NASA Exoplanet Archive. The dominant contribution to the total blend frequency, as is often the case for \kepler\ candidates, was from false positives involving an unseen companion to the target star that
is in a wide orbit and is transited by a larger planet (HTP). On the other hand, the expected rate of occurrence of transiting planets of the size of \koicurb, based again on the available sample of KOIs from the NASA Exoplanet Archive, is $9.97 \times 10^{-4}$. The `odds ratio' is therefore $9.97 \times 10^{-4} / 2.35 \times 10^{-6} \approx 424$.  Consistent with previous applications of \blender, we adopted a threshold for validation of $1/(1/99.73\%-1) \approx 370$, corresponding to a $3\,\sigma$ significance. Any odds ratio above this has an even greater confidence level. This means that \koicurb\ is firmly validated at the 99.76\% level.


\section{Habitability and Composition of \koicurb}\label{s:habitabilityandcomposition}
Here we examine the probability that \koicurb\ orbits in the habitable zone of its star, consider the effects of the unknown eccentricity on the habitability, and estimate the likelihood that it is rocky. Of chief interest is whether the insolation flux experienced by \koicurb\ would permit water to exist in liquid form on its surface. Secondarily, we wished to determine the likelihood that it possesses a rocky surface upon which liquid water could pool. 

\subsection{Habitability}\label{ss:habitability}
Following the statistical approach taken by \citet{torres2015}, we took the insolation flux levels defined in \citet{kopparapu2013} for recent Venus and for early Mars to be the inner and outer boundaries of the wide or optimistic habitable zone, and the insolation flux levels for the runaway greenhouse and a maximum greenhouse as the inner and outer boundaries for the narrow or conservative habitable zone. This was somewhat more conservative, as \citet{torres2015} took the inner boundary of the habitable zone to correspond to that for dry desert worlds from \citet{zsom2013}, which permits habitable worlds as close as 0.38\,AU from a solar-like host star if the albedo is high and the relative humidity is low ($\sim$1\%). 

\begin{figure}[ht]
\resizebox{\hsize}{!}{\includegraphics{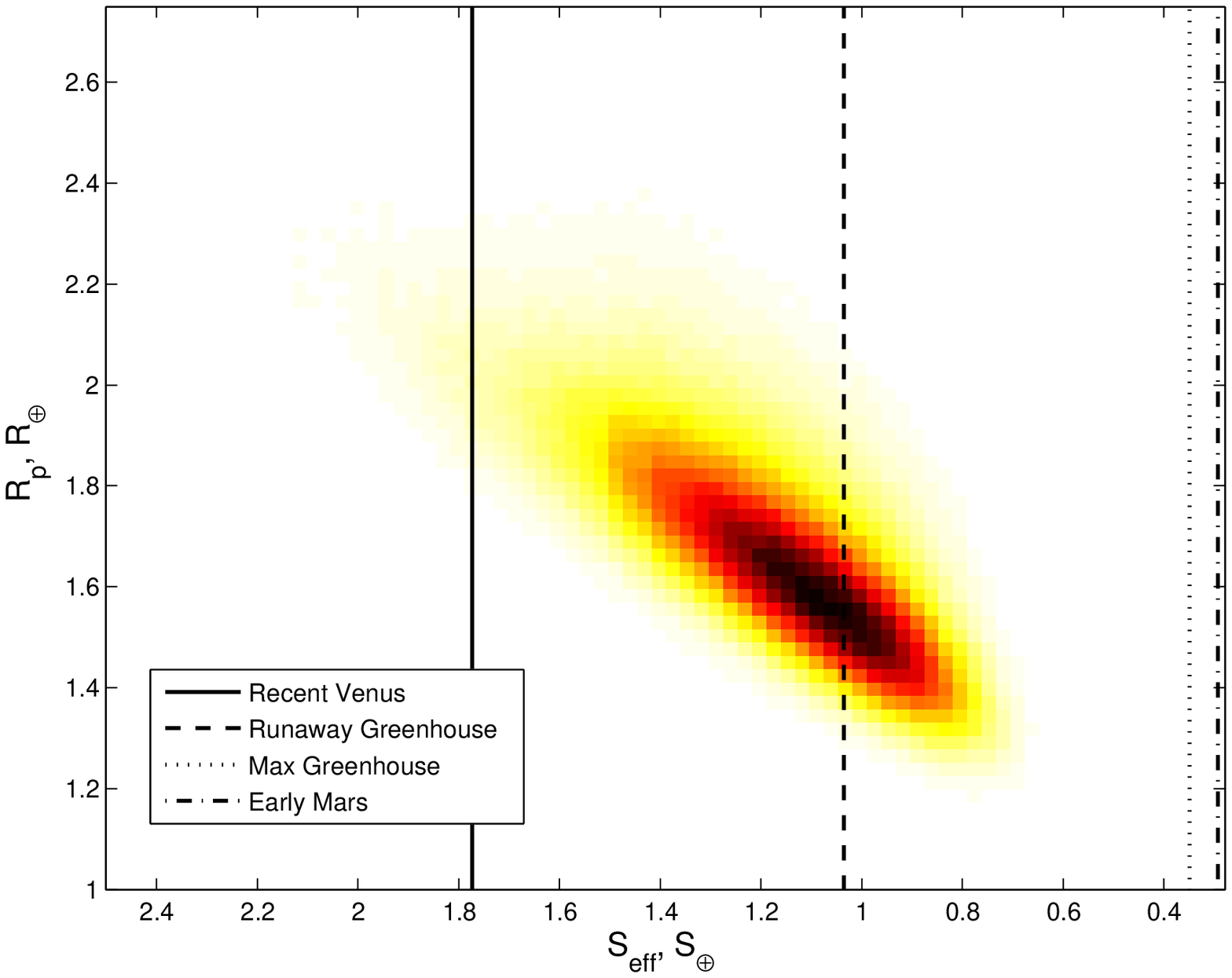}}
\caption{
The joint density of the planetary radius, $\rpl$, of \koicurb\ and its insolation flux, \seff, in Earth insolation units, displayed as a false color image. The vertical lines show the insolation flux values for the modes of the MCMC realizations for the SpecMatch stellar parameters presented in Section~\ref{s:hostStarProperties} as developed by \citet{kopparapu2013}: the optimistic or wide habitable zone is bounded on the inside by recent Venus (solid line) and on the outside by early Mars (dashed dotted line), and the conservative or narrow habitable zone is bounded on the inside by the runaway greenhouse effect (dashed line) and on the outside by the maximum greenhouse effect (dotted line).
 \label{fig:RpVsShistogram}}
\end{figure}

The 900,000 MCMC chains for the stellar parameters discussed in Section~\ref{s:hostStarProperties} were convolved with realizations of the measured orbital period distribution to construct a population of insolation flux levels relative to Earth's insolation, calculated as
\begin{equation}
\frac{\seff}{S_\oplus} =  a^{-2}\left(\frac{\teff}{T_\sun} \right)^4\,\left(\frac{R_\ast}{\rsun}\right)^2,
\end{equation}
where $T_\sun$ = 5778 K is the solar temperature, \teff\ is the effective temperature of the star, $a$ is the mean separation between the star and the planet in AU, and $R_\ast$ is the radius of the star. For each realized effective temperature value, the insolation flux values corresponding to the inner boundaries of the optimistic and conservative habitable zones were calculated as per \citet{kopparapu2013} and the sample was marked as habitable if the insolation was less than that for recent Venus/runaway greenhouse. Figure~\ref{fig:RpVsShistogram} shows a false color image of the joint distribution between insolation and planetary radius for all 900,000 realizations, where the planetary radius samples were generated by combining the stellar radius samples with a realization for the transit depth, or ratio of the planetary radius to that of the parent star, $\rpl/R_\ast$. Marginalizing the joint distribution over \seff, we find the likelihood that \koicurb\ falls within the wide or optimistic habitable zone is \HZprobabilitySpecMatchRecentVenus\%, while the likelihood is \HZprobabilitySpecMatchRunawayGreenhouse\% that it falls within the narrow or conservative habitable zone. 

Given that the SpecMatch results are somewhat conservative with respect to the surface gravity, we repeated the calculations above for the SPC stellar parameters obtained for the HIRES spectrum as a means by which to explore the reasonable non-conservative limits. The SPC parameters yielded realizations that are within the wide and the narrow habitable zone \HZprobabilitySPCrecentVenus\% and \HZprobabilitySPCrunawayGreenhouse\% of the time, respectively.

\subsection{Effect of Eccentricity on Habitability}\label{ss:eccentricity}
The eccentricity of the orbit of \koicurb\ is poorly constrained by the photometric measurements, at best. With an orbital period of \KOIperiodshort\ days, and a mass of $\sim$5 \mearth (see Section~\ref{ss:composition}), \koicurb\ should exhibit a reflex velocity of $\sim$45\,cm\,s$^{-1}$. As the host star is relatively faint for such work (\Kp\ =  \koicurCCkicmag), there is little chance of measuring its orbit through radial velocity observations in the near future.  

\begin{figure}[ht]
\resizebox{\hsize}{!}{\includegraphics{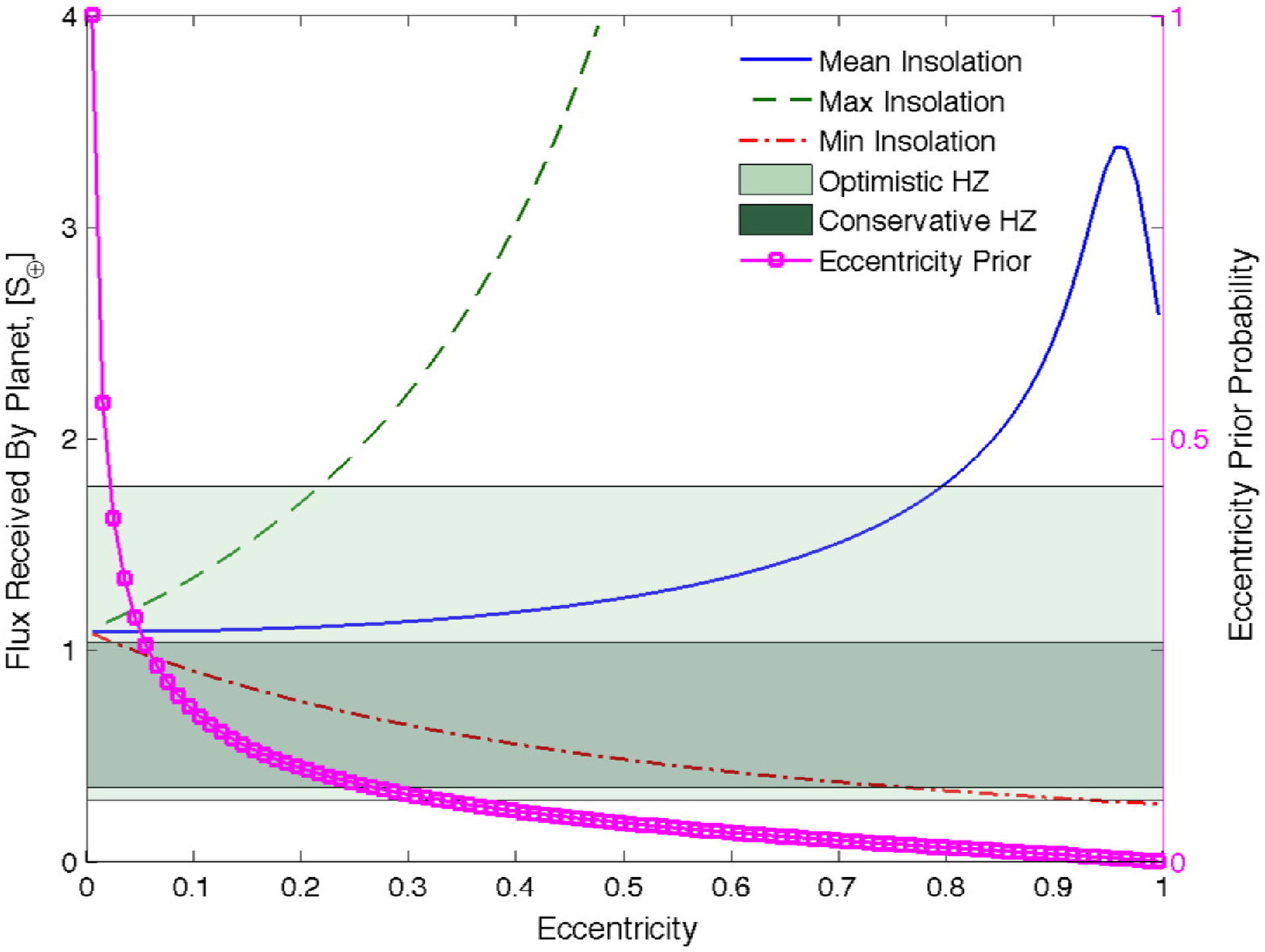}}
\caption{
The insolation flux, \seff, in Earth insolation units received by the planet \koicurb\ as a function of orbital eccentricity, along with a prior distribution for eccentricity for transiting planets as per \citet{roweMultis2014}. The mean, max, and min insolation flux values are indicated by the solid blue, the dashed green and the dashed-dotted red curves, respectively. The optimistic and conservative habitable zones, according to \citet{kopparapu2013}, are indicated by the light and dark green shaded regions, respectively. The eccentricity prior is indicated by the solid magenta curve with the square symbols.
 \label{fig:eccentricity}}
\end{figure}

To investigate the effect of the uncertainties of the eccentricity of the orbit on the habitability of \koicurb, we examined the variation of the mean insolation flux as a function of eccentricity. Figure~\ref{fig:eccentricity} shows the mean, maximum and minimum insolation flux for realizations of an orbit at the semi major axis and orbital period of \koicurb\ as a function of the eccentricity from 0 to 1. While the maximum insolation flux exceeds 4\,\Searth\ for $e > 0.475$, the mean flux does not exceed the maximum permitted by the optimistic habitable zone (recent Venus) until $e$ exceeds 0.8. The  probability density for eccentricity developed by \citet{roweMultis2014} based on the population of \kepler\ multiple transiting planets is also shown on this figure, illustrating the fact that transiting planets exhibit very low eccentricities in general. \citet{roweMultis2014} modeled the eccentricity distribution as a beta distribution $\beta\left(x,y\right)$ with $x$ = 0.4497 and $y$ = 1.7938.  The eccentricity distribution-weighted mean insolation flux for \koicurb\ is 1.17 \Searth, only 9\% higher than that for $e$ = 0. Thus we conclude that the uncertainties in the eccentricity of \koicurb\ do not significantly affect the habitability of its orbit.

\subsection{Composition}\label{ss:composition}
Because \koicurb\ does not have a mass measurement, estimating the probability that it is rocky necessarily requires the application of statistical mass-radius (M--R) relationships.  One such relation is provided by \citet{weissAndMarcy2014}, where the mass of the planet is a deterministic function of its radius.  Alternatively, the analysis of \citet{wolfgang2015} yields a probabilistic M--R relation that allows a range of masses to exist at a given radius, as we would expect if there is a distribution of compositions among the population.  Either can be used to produce $\left(\mpl,\rpl\right)$ pairs that can then be compared to the theoretical M--R curves of \citet{fortneyMarleyBarnes2007}: If the radius is smaller at the assigned mass than that for a 100\% silicate planet, as determined by the 100\% rock mass fraction in the rock-iron composition curve (Equation (8) of that paper), then that $\left(\mpl,\rpl\right)$ pair is consistent with being rocky. We applied both the deterministic  M--R relationship of \citet{weissAndMarcy2014} and the probabilistic formulation of \citet{wolfgang2015} to investigate the likelihood that \koicurb\ is rocky.

\koicurb's radius is actually a distribution which incorporates measurement uncertainty reflected in the posterior distribution of the MCMC realizations. To calculate the probability that it is rocky, we drew a set of radii from this distribution, applied the M--R relations to each to obtain mass estimates, and measured the fraction of draws which were consistent with a rocky planet as defined above.  Here the radius distribution was determined by both the $\slfrac{\rpl}{R_\star}$ posterior produced by the lightcurve fitting of Section~\ref{s:planet properties}, and the stellar radius posterior of Section~\ref{s:hostStarProperties} and Figure ~\ref{fig:stprop_mcmc}; drawing randomly from each of these sets of posterior samples and multiplying them then gave the set of planet radius posterior samples used for this analysis.

Because the \citet{weissAndMarcy2014} M--R relation is a deterministic one, applying it to this problem is straightforward.  However, it does not explicitly capture the intrinsic variability in planet masses that we both expect and observe, so the rocky probabilities that it produces may be inaccurate and biased.  To improve on this, we computed the posterior predictive distribution for \koicurb's mass given the probabilistic relationship of \citet{wolfgang2015}.  This involved drawing posterior samples from the M--R relation hyperparameters (i.e. the power law constant, index, and the mass dispersion around this power law) that were produced by evaluating the hierarchical Bayesian model in this sample \citep[see][for more details]{wolfgang2015}.  This draw defined the M--R relation which then yielded the mass corresponding to the drawn radius.  The resulting posterior predictive distribution therefore accounts for both the intrinsic, expected dispersion in planet compositions in this small-radius range and for the uncertainties in the parameter values themselves.  As a result, the \citet{wolfgang2015} approach most accurately incorporates the degree of our uncertainty in this radius-to-mass ``conversion."

Specifically, we took the deterministic mass--density relation from Equations (1)--(3) in \citet{weissAndMarcy2014}:
\begin{align}
\mpl = \min{\left\lbrace \frac{2.43+3.39\,\rpl} {5.513} \, {\rpl}^3, \; 2.69\,\rpl^{0.93}\right\rbrace},
\end{align}
where \mpl\ and \rpl\ are the planetary mass and radius, respectively, in Earth units.  

The probabilistic mass-radius relation of \citet{wolfgang2015} has the form:
\begin{equation}
\mpl \sim N\left(\mu= 2.7 \,\rpl^{\,\,1.3} ;\sigma=1.9\right),
\end{equation}
where $N\left(\mu;\sigma\right)$ denotes a normal (Gaussian) distribution with mean, $\mu$, and standard deviation, $\sigma$, and where \rpl\ and \mpl\ are in Earth units.

The radius of a pure rock composition planet was calculated as per Eq.~8 in \citet{fortneyMarleyBarnes2007} for the estimated planetary mass:
\newcommand{\frock}{\ensuremath{f_{\rm rock}}}
\begin{equation}
\begin{split}
R_\mathrm{rock} = & \left(0.0592\,\frock+0.0975\right)\,\left(\log{\mpl}\right)^2 + \\
& \left(0.2337\,\frock+0.4938\right)\,\log{\mpl}+ \\
& \left(0.3102\,\frock+0.7932\right),
\end{split}
\end{equation}
where \frock\ is the rock mass fraction (set to 1 for this analysis).

\begin{figure}[ht]
\resizebox{\hsize}{!}{\includegraphics{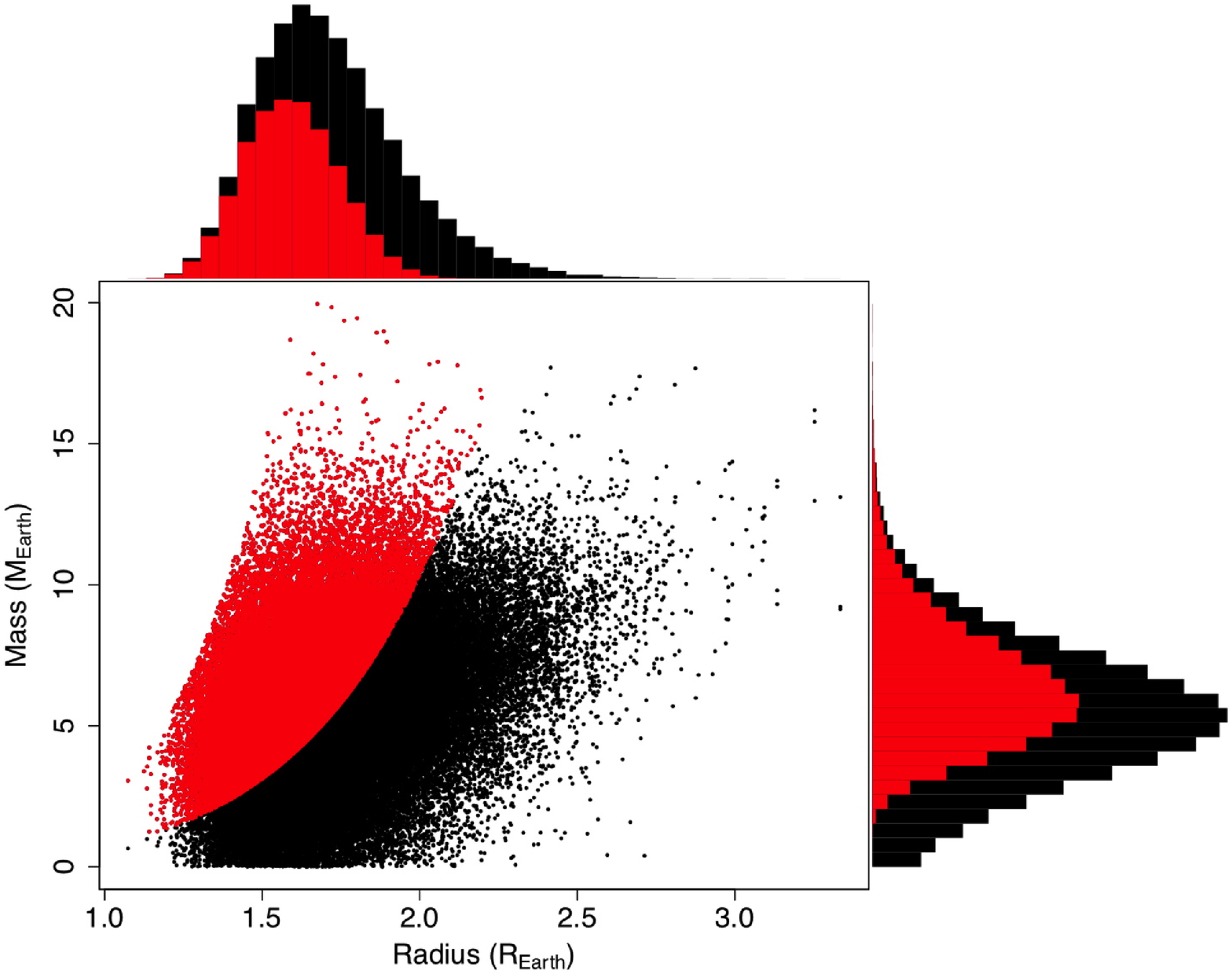}}
\caption{
The joint density of the planetary radius, \rpl, of \koicurb\ and its mass, \mpl, in Earth units, displayed as a scatter plot with marginal histograms for each quantity for the SpecMatch derived parameters. The planetary mass realizations were generated using a hierarchical Bayesian model as per \citet{wolfgang2015} which explicitly models the planetary mass as a distribution for a given planetary radius. The data and histograms are colored red for those realizations that are denser than predictions for 100\% silicate planets as per \citet{fortneyMarleyBarnes2007}, and black for those realizations that are less dense and likely to be non-rocky. The fraction of realizations allows us to estimate the probability to be \rockyProbabilitySpecMatchWolfgang\% that \koicurb\ is rocky for this set of stellar and planetary parameters.
 \label{fig:hbmspecmatch}}
\end{figure}

We computed the probabilities that \koicurb\ is rocky for both the conservative SpecMatch stellar parameters and for the SPC results.  Applying the M--R relation of \citet{weissAndMarcy2014} gave rocky probabilities of \rockyProbabilitySpecMatchWeiss\% and \rockyProbabilitySPCWeiss\%, respectively.  The posterior predictive distribution displayed in figure~\ref{fig:hbmspecmatch} obtained using the probabilistic M--R relation from \citet{wolfgang2015} gave \rockyProbabilitySpecMatchWolfgang\% and \rockyProbabilitySPCWolfgang\%, respectively.  The good agreement between the results for both M--R relationships lends confidence in the results and we estimate the likelihood that \koicurb\ is rocky as between that for the SpecMatch and the SPC parameters, namely, between \rockyProbabilitySpecMatchWolfgang\% and \rockyProbabilitySPCWolfgang\%. 

We note that it is unlikely that \koicurb\ has an Earth-like composition: 
we computed the density of Earth-like composition planets for the SpecMatch realizations according to \citet[][i.e., a rock-iron mixing ratio of 2/3 in Eq.~8 therein]{fortneyMarleyBarnes2007} and found that only \earthCompositionProbabilitySpecMatchWeiss\% and \earthCompositionProbabilitySpecMatchWolfgang\% of the \citet{weissAndMarcy2014} and of the \citet{wolfgang2015} realizations, respectively, were at least as dense as that predicted for Earth-composition planets.

\subsection{History of the Habitability of \koicurb}\label{ss:historyofhabitability}
Given the similarities between the Sun and \koicurCCkic\ with respect to mass and effective temperature, and the fact that the orbital distance of \koicurb\ is essentially 1\,AU, it is interesting to consider the habitability of this planet over the history of this planetary system. 
It is important to note that statements comparing the age of \koicurCCkic\ to that of the Sun are tentative, given that the estimated age for \koicurCCkic\ is heavily model dependent and the ~2\,Gyr uncertainty estimated in Section~\ref{s:hostStarProperties} does not take into account variations of poorly constrained model physics such as convection (mixing length parameter) and microscopic diffusion. In this section, we use our best estimate for the stellar age to consider a case study of the evolution of the habitable zone over the main-sequence lifetime for stars such as \koicurCCkic\ and our own Sun.

We traced the evolution of the effective temperature and the luminosity of \koicurCCkic\ using the Dartmouth stellar evolution models \citep{Dotter08}\footnote{http://stellar.dartmouth.edu/models/} appropriate to its mass (\KOImstarshort\ \msun) and metallicity (\feh\ = \KOIfehshort). The evolving effective temperature allowed us to determine the inner and outer radii of the conservative and optimistic habitable zone per \citet{kopparapu2013} as a function of time. The luminosity history obtained from the Dartmouth models allowed us to estimate the radiation received by this super Earth over the main-sequence lifetime of its star.

\begin{figure}[ht]
\resizebox{\hsize}{!}{\includegraphics{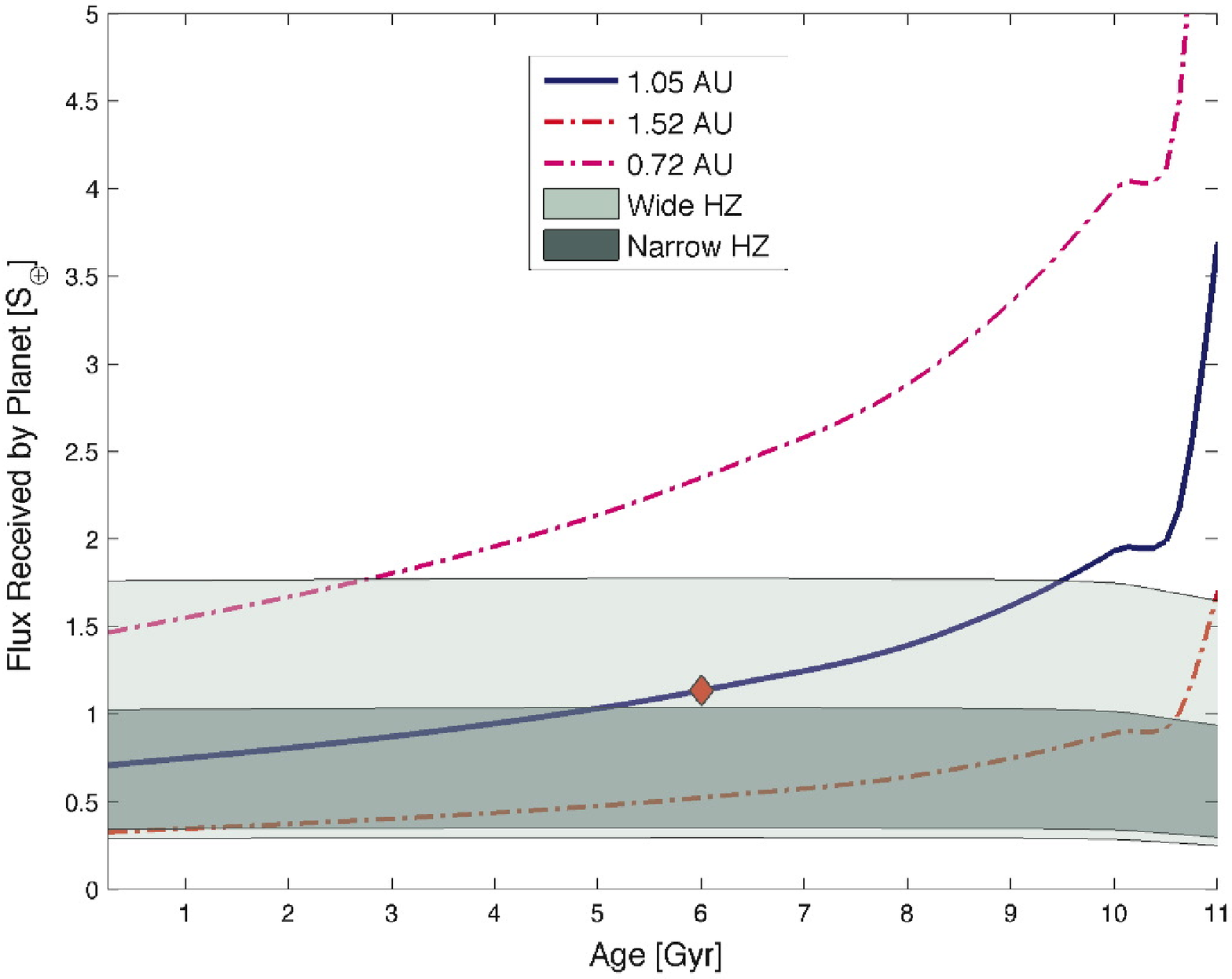}}
\caption{
A history of the habitable zone of \koicurCCkic\ as delineated by the amount of flux received by a planet at the orbital distance of \koicurb\ (1.05\,AU -- essentially the same as that of the Earth) over the main-sequence lifetime of its star. This figure plots the radiation received by \koicurb\ from 0.25\,Gyr to 11\,Gyr as a thick blue curve. The optimistic and conservative habitable zones, according to \citet{kopparapu2013}, are indicated by the light and dark green shaded regions, respectively. The planet is currently experiencing radiation at a level 10\% above that received by Earth. However, over the first 5.2\,Gyr of its history, \koicurb\ was in the conservative habitable zone, receiving less than 1.0369 of the flux received by contemporary Earth. The red and purple stippled curves represent the flux received by planets in Mars' and Venus' orbits (1.52 and 0.72\,AU), respectively.
 \label{fig:habzone}}
\end{figure}

At $\sim$\KOIageshort\,Gyr, \koicurCCkic\ and its planet are about 1.5\,Gyr older than the solar system, and the host star's radius is $\sim$10\% larger than that of the Sun. This planet bathes in a stream of radiation in its orbit at a level only $\sim$10\% higher than that experienced by Earth today. However, since the mass of the host star is only $\sim$4\% greater than that of the Sun, \koicurb\ spent the first $\sim$5\,Gyr in the conservative habitable zone according to \citet{kopparapu2013}, as illustrated in Figure~\ref{fig:habzone}. This planet will remain in the optimistic habitable zone for another $\sim$3.5\,Gyr before the star leaves the main-sequence and expands rapidly in the ascending red giant branch phase.

If this exoplanet system is a future version of an analog to our own solar system, with additional planets in the orbits corresponding to those of Mars and Venus, their insolation history would be very similar to their namesakes in the solar system. An exo-Venus would have spent only the first $\sim$3\,Gyr in the optimistic habitable zone before exiting it for good. In contrast, an exo-Mars would be located in the conservative habitable zone during the entire $\sim$11\,Gyr spent by the host star on the main-sequence.

At present, \koicurb\ is the only known planet orbiting this star. Assuming that the dispersion in orbital inclination angles is the same in the \koicurCCkic\ system as it is for the solar system ($\sim$1\fdg9), \kepler\ would have had only a $\sim$10\% chance of observing a transiting exo-Venus in this system as well as \koicurb. The probability of observing a few transits of an exo-Mars in addition to those of \koicurb\ is $\sim$6\%, but it would have to have been a rather large planet in order for its transits to be readily identifiable in the light curve by eye since it would have provided fewer than three transits over the course of the \kepler\ Mission. No such transits have been seen. \kepler\  had only a $\sim$2\% chance of seeing transits of both an exo-Venus and an exo-Mars in addition to those of the super Earth at \KOIashort\,AU. Until such time as other planets are discovered by other exoplanet surveys, or a SETI observation of this planetary system reveals signatures of extraterrestrial technologies, we are left to speculate on the fate of an ancient civilization that may have developed first on \koicurb\ (or on a moon orbiting it). For example, it  may have subsequently migrated to an as-yet-undetected outer planet to escape the inevitable loss of most of the intrinsic water inventory after the moist runaway greenhouse effect took hold at \koicurb's orbital distance approximately 800 Myr ago.

\section{Conclusions}\label{s:conclusions}
The identification of a \KOIRpshort-\rearth\ planet in the habitable zone of a G2 star at a mean separation of \KOIashort\,AU represents the closest analog to the Earth--Sun system discovered in the \kepler\ data set to date with respect to the orbit and spectral type of the host star. Moreover, the small radius of this planet provides a reasonable chance between \rockyProbabilitySpecMatchWolfgang\% and \rockyProbabilitySPCWolfgang\% that it is rocky, although in this case it is unlikely to have an iron core of significant mass.  

Given our current best estimates the host star is ~10\% larger and ~1.5\,Gyr older than the Sun; hence, \koicurb\ may represent the future environment that the Earth will experience as our own Sun evolves toward the red giant phase. 
Though it is currently just beyond the moist runaway greenhouse level for the incident radiation, this planet is still within the optimistic habitable zone as per \citet{kopparapu2013}, and will remain so for another $\sim$3\,Gyr. However, \koicurb\ likely spent the first $\sim$5\,Gyr in the conservative habitable zone, below the radiation level experienced by Earth today.

The paucity of small planets in or near the habitable zone of solar-like stars, particularly G stars, makes the determination of the intrinsic frequency of Earth-size planets in the habitable zone of Sun-like stars difficult to determine robustly without resorting to extrapolation. \koicurb\ represents an important addition to the list of small planets in or near the habitable zone for its proximity to a G2 star. We are hopeful that this is only the first of many such planetary systems yet to be discovered in the fullness of time.
 
\acknowledgements 
\kepler\ was competitively selected as the 10th Discovery mission. Funding for this mission is provided by NASA's Science
Mission Directorate.  The authors acknowledge the efforts of the \kepler\ Mission team in obtaining the light curve data and data validation products used in this publication. These data were generated by the \kepler\ Mission science pipeline through the efforts of the \kepler\ Science Operations Center and Science Office. The \kepler\ Mission is led by the project office at NASA Ames Research Center. Ball Aerospace built the \kepler\ photometer and spacecraft which is operated by the mission operations center at LASP. The \kepler\ light curves are archived at the Mikulski Archive for Space Telescopes and the Data Validation products are archived at the NASA Exoplanet Science Institute. Some of the data presented herein were obtained at the W. M. Keck Observatory, which is operated as a scientific partnership of the California Institute of Technology, the University of California, and NASA. The Keck Observatory was made possible by the generous financial support of the W. M. Keck Foundation.
The authors wish to recognize and acknowledge the very significant cultural role and reverence that the summit of Mauna Kea has always had within the indigenous Hawaiian community.  We are most fortunate to have the opportunity to conduct observations from this mountain. 
M.~E. acknowledges support by NASA under grant NNX14AB86G issued 
through the \kepler\ Participating Scientist Program.
D.~W.~L. acknowledges partial support from NASA's \kepler\ Mission under Cooperative Agreements NNX13AB58A with the Smithsonian Astrophysical Observatory. 
G.~T. acknowledges support for this work from NASA under grant NNX14AB83G (\kepler\  Participating Scientist Program).
D.~H. acknowledges support by the Australian Research Council's Discovery Projects funding scheme (project number DE140101364) and support by the National Aeronautics and Space Administration under Grant NNX14AB92G issued through the \kepler\ Participating Scientist Program.
A.~W.'s funding was provided by the National Science Foundation Graduate Research Fellowship under Grant No. 0809125, and by the UC Santa Cruz Graduate Division's Eugene Cota-Robles Fellowship. 
J.~M.~J. acknowledges 20 years of unwavering support from Ren\'ee M. Schell and excellent editorial suggestions for this paper. We dedicate this discovery to the memory of \kepler's late Deputy PI, Dr. David G. Koch, who represented the epitome of grace, intellect and civility, and who would have been deliriously delighted by the discovery of this planet.

{\it Facilities:} Kepler.

\appendix
\section{A Computationally Efficient Statistical Bootstrap Test for Transiting Planets}\label{app:bootstrap}

To search for transit signatures, TPS employs a bank of wavelet-based matched filters that form a grid on a three-dimensional parameter space of transit duration, orbital period, and phase \citep{jenkins2002,jenkinsTPS2010}.  A detection statistic is calculated for each template and compared to a threshold value of $\eta = 7.1\, \sigma$.  Since TPS searches the light curve by folding the single-transit detection statistics at each trial orbital period, the detection statistic is referred to as a multiple event statistic, or $MES$. Detections in TPS are made under the assumption that the pre-whitening filter applied to the light curve yields a time series whose underlying noise process is stationary, white, Gaussian, and uncorrelated.  When the pre-whitened noise deviates from these assumptions, the detection thresholds are invalid and the false alarm probability associated with such a detection may be significantly higher than that for a signal embedded in white Gaussian noise.  The Statistical Bootstrap Test, or the Bootstrap, is a way of building the distribution of the null statistics from the data so that the false alarm probability can be calculated for each TCE based on the observed distribution of the out-of-transit statistics.  

Consider a TCE exhibiting $p$ transits of a given duration within its light curve. In this analysis, the light curve is viewed as one realization of a stochastic process. Further, consider the collection of all $p$-transit detection statistics that could be generated if we had access to an infinite number of such realizations. To approximate this distribution we formulate the single event statistic time series for the light curve and exclude points in transit (plus some padding) for the given TCE. Bootstrap statistics are then generated by drawing at random from the single event statistics $p$ times with replacement to formulate the $p$-transit statistic (i.e., the $MES$). Since the single event statistics encapsulate the effects of local correlations in the background noise process on the detectability of transits, so does each individual bootstrap statistic. So long as the orbital period is sufficiently long (generally longer than several hours), the single event statistics are uncorrelated and the $MES$ can be considered to be formed by $p$ independent random deviates from the distribution of null single event statistics.

\citet{jenkinsetal2002} formulated a bootstrap test for establishing the confidence level in planetary transit signatures identified in transit photometry in white, but possibly non-Gaussian noise. \citet{jenkins2002} extended this approach to the case of non-white noise. In both cases, the bootstrap false alarm rate as a function of  the $MES$ of the detected transit signature was estimated by explicitly generating individual bootstrap statistics directly from the set of out-of-transit data. This direct bootstrap sampling approach can become extremely computationally intensive as the number of transits for a given TCE grows beyond 15. The number of individual bootstrap statistics that can be formed from the $m$ out-of-transit cadences of a light curve and the $p$ transits is $m^p$, which is $\sim2.9\times10^{48}$ statistics for 10 transits and 4 years of \kepler\ data. An alternative, computationally efficient method can be implemented by formulating the bootstrap distribution in terms of the probability density function (PDF) of the single event detection statistics. The distribution for the $MES$ as a function of threshold can then be obtained from the distribution of single event statistics treated as a bivariate random process.

The $MES$, $Z$,  can be expressed as
\begin{equation}
Z =\left . \sum_{i \in \mathcal{S}} \mathbb{C}(i) \middle\slash \sqrt{ \sum_{i \in \mathcal{S}} \mathbb{N}(i) }\right .,
\end{equation}  
where $\mathcal{S}$ is the set of transit times that a single period and epoch pair select out, $\mathbb{C}(i)$ is the correlation time series formed by correlating the whitened data to a whitened transit signal template with a transit centered at the $i$th timestep in the set $\mathcal{S}$, and $\mathbb{N}(i)$ is the template normalization time series. The square root of the normalization time series, $\sqrt{\mathbb{N}(i)}$, is the expected value of the $MES$ or S/N for the reference transit pulse.\footnote{The inverse of $\sqrt{\mathbb{N}(i)}$ can be interpreted as the effective white Gaussian noise ``seen'' by the reference transit and is the definition for the combined differential photometric precision (CDPP) reported for the \kepler\ light curves at 3, 6, and 12 hour durations.}

If the observation noise process underlying the light curve is well modeled as a possibly non-white, possibly non-stationary Gaussian noise process, then the single event statistics will be zero-mean, unit-variance Gaussian random deviates. The false alarm rate of the transit detector would then be described by the complementary distribution for a zero-mean, unit-variance Gaussian distribution:
\begin{equation}
\bar{F}_Z\left(Z\right) = \frac{1}{2} \erfc\left(Z\middle\slash\sqrt{2}\right),
\end{equation}
where $\erfc\left(\cdot\right)$ is the standard complementary error function. If the power spectral density of the noise process is not perfectly captured by the whitener in TPS, then the null statistics will not be zero-mean, unit-variance Gaussian deviates. A bootstrap analysis allows us to obtain a data-driven approximation of the actual distribution of the null statistics, rather than relying on the assumption that the pre-whitener is perfect.

The random variable $Z$ is a function of the random variables corresponding to the correlation and normalization terms in the single event statistic time series, $\mathbb{C}_p = \sum_{i \in \mathcal{S}} \mathbb{C}(i)$, and  $\mathbb{N}_p = \sum_{i \in \mathcal{S}} \mathbb{N}(i)$. The joint density of $\mathbb{C}_p$ and  $\mathbb{N}_p$ can be determined from the joint density of the single event statistic components $\mathbb{C}$ and  $\mathbb{N}$ as
\begin{equation}
f_{\mathbb{C}_p,\mathbb{N}_p}\left(\mathbb{C}_p,\mathbb{N}_p\right) = f_{\mathbb{C},\mathbb{N}}\left(\mathbb{C},\mathbb{N}\right) \ast f_{\mathbb{C},\mathbb{N}}\left(\mathbb{C},\mathbb{N}\right) \ast \ldots\ast f_{\mathbb{C},\mathbb{N}}\left(\mathbb{C},\mathbb{N}\right),\label{eq:convolution}
\end{equation}
where `$\ast$' is the convolution operator and the convolution is performed $p$ times. This follows from the fact that the bootstrap samples are constructed from \emph{independent} draws from the set of null (single event) statistics with replacement.\footnote{In this implementation, we choose to assume that the null statistics are governed by a single distribution. If this is not the case (for example, if the null statistic densities vary from quarter to quarter) then Eq.~\ref{eq:convolution} could be modified to account for the disparity in the relevant single event statistics in a straightforward  manner.} Given that convolution in the time/spatial domain corresponds to multiplication in the Fourier domain, Eq. \ref{eq:convolution} can be represented in the Fourier domain as
\begin{equation}
\Phi_{\mathbb{C}_p,\mathbb{N}_p} = \Phi_{\mathbb{C},\mathbb{N}} \cdot \Phi_{\mathbb{C},\mathbb{N}} \cdot \ldots\cdot \Phi_{\mathbb{C},\mathbb{N}} = \Phi_{\mathbb{C},\mathbb{N}}^p,\label{eq:fourier}
\end{equation}
where $\Phi_{\mathbb{C},\mathbb{N}}=\mathscr{F}\left\lbrace f_{\mathbb{C},\mathbb{N}}\right\rbrace$ is the  Fourier transform of the joint density function $f_{\mathbb{C},\mathbb{N}}$. Here, the arguments of the Fourier transforms of the density functions have been suppressed for clarity. The use of 2D fast Fourier transforms results in a highly tractable algorithm from a computational point of view.

The implementation of Eq.~\ref{eq:fourier} requires that a 2D histogram be constructed for the $\left\lbrace \mathbb{C},\mathbb{N}\right\rbrace$ pairs over the set of null statistics. Care must be taken to manage the size  of the histogram to avoid spatial aliasing as the use of fast Fourier transforms corresponds to circular convolution. We chose to formulate the 2D grid to allow for as high as $p=8$ transits. The intervals covered by the realizations (i.e., the support) for each of $\mathbb{C}$ and $\mathbb{N}$ were sampled with 256 bins, and centered  in a 4096 by 4096 array. When $p>8$, it is necessary to implement Eq.~\ref{eq:fourier} iteratively in stages, after each of which the characteristic function is transformed back into the spatial domain, then bin-averaged by a factor of 2 and padded back out to the original array size. Care must also be taken to manage knowledge of the zero-point of the histogram in light of the circular convolution, as it shifts by 1/2 sample with each convolution operation in each dimension. 

Once the $p$-transit 2D density function $f_{\mathbb{C}_p,\mathbb{N}_p}\left(\mathbb{C}_p,\mathbb{N}_p\right)$ is obtained, it can be ``collapsed'' into the sought-after one-dimensional density $f_Z\left(Z\right)$ by mapping the sample density for each cell with center coordinates $\left \lbrace \mathbb{C}_i,\mathbb{N}_j\right\rbrace$ to the corresponding coordinate $Z_{i,j} = \mathbb{C}_i/\sqrt{\mathbb{N}_j}$, and formulating a histogram with a resolution of, say, 0.1 $\sigma$ in $Z$ by  summing the resulting densities that map into the same bins in $Z$. Due to the use of FFTs, the precision of the resulting density function is limited to the floating point precision of the variables and computations, which is $\sim2.2\times10^{-16}$. For small $p$, the density may not reach to the limiting numerical precision because of small number statistics, and for large $p$, roundoff errors can accumulate below about $10^{-14}$. The results can be extrapolated to high $MES$ values by fitting the mean, $\mu$, and standard deviation, $\sigma$, of a Gaussian distribution to the empirical distribution in the region $10^{-4}\le\bar{F}_Z\le10^{-13}$ using the standard complementary error function: $\bar{F}_Z\left(Z\right) = 0.5 \erfc\left(\left(Z-\mu\right)/\sqrt{2}\, \sigma\right)$.

In order to model the use of $\chi$-square vetoes \citep{seaderVetos2013}, we pre-filtered the single event statistic time series to remove the three most positive peaks and their ``shoulders'' down to 2 $\sigma$. The three most negative peaks were also handled in a similar fashion to avoid biasing the mean of the null statistics in a negative direction. We also identified and removed points with a density of zero-crossings that fell below 1/4 that of the median zero-crossing density. This step removed single event statistics in regions where the correlation term experienced strong excursions from zero due to unmitigated sudden pixel sensitivity dropouts and thermal transients near monthly and quarterly boundaries. Typically, more than 99\% of the original out-of-transit single event statistics were retained by these pre-filters for the 16,285 TCEs resulting from the Q1--Q16 planet search documented in \citet{tenenbaum2014}.

\vspace{0.25cm}
 

\end{document}